\documentclass[twocolumn]{aastex62}
\usepackage{graphics,epsf}
\usepackage{amsmath}                
\usepackage{amsfonts}               
\usepackage{amssymb}                
\usepackage{epsfig}                 
\usepackage{graphicx}
\usepackage{float}
\usepackage{color}
\usepackage[para,online,flushleft]{threeparttable}

\newcommand{\cm}{{~\rm cm}}
\newcommand{\km}{{~\rm km}}
\newcommand{\s}{{~\rm s}}

\newcommand{\g}{{~\rm g}}

\newcommand{\K}{{~\rm K}}
\newcommand{\erg}{{~\rm erg}}

\newcommand{\AU}{{~\rm AU}}

\newcommand{\days}{{~\rm days}}

\begin{document}

\title{Simulating the inflation of bubbles by late jets in core collapse supernova ejecta} 


\author{Muhammad Akashi}
\affiliation{Department of Physics, Technion, Haifa, 3200003, Israel; akashi@physics.technion.ac.il; soker@physics.technion.ac.il}
\affiliation{Kinneret College on the Sea of Galilee, Samakh 15132, Israel}

\author[0000-0003-0375-8987]{Noam Soker}
\affiliation{Department of Physics, Technion, Haifa, 3200003, Israel; akashi@physics.technion.ac.il; soker@physics.technion.ac.il}
\affiliation{Guangdong Technion Israel Institute of Technology, Shantou 515069, Guangdong Province, China}

\begin{abstract}
We conducted three-dimensional hydrodynamical simulations to study the interaction of two late opposite jets with the ejecta of a core collapse supernova (CCSN), and study the bipolar structure that results from this interaction as the jets inflate hot-low-density bubbles. The newly born central object, a neutron star (NS; or a black hole), launches these jets at about 50 to 100 days after explosion. The bubbles cross the photosphere in the polar directions at much earlier times than the regions at the same radii near the equatorial plane. The hot bubbles releases more radiation and the photosphere recedes more rapidly in the tenuous bubble. Our results strengthen earlier claims that were based on toy models that such an interaction might lead to a late peak in the light curve, and that an equatorial observer might see a rapid drop in the light curve.
Our results  might  have implications to much earlier jets that explode the star, either jets that the newly born NS launches in a CCSN, or jets that a NS companion that merges with the core of a massive star launches in a common envelope jets supernova (CEJSN) event. Our results add indirect support to the CEJSN scenario for fast blue optical transients, e.g., AT2018cow, ZTF18abvkwla, and CSS161010.
\end{abstract}


\section{Introduction}
\label{sec:intro}
    
There are several types of observations that point directly and indirectly at the possible role of jets in, at least some, core collapse supernovae (CCSNe). One observation is a bipolar morphology of the $^{56}$Ni in the Type II-P CCSN SN~2016X (ASASSN-16at;
\citealt{Boseetal2019}). Jets that drive CCSNe form bipolar morphological features (e.g., \citealt{Orlandoetal2016, BearSoker2018}), and therefore might account for this $^{56}$Ni morphology. Other observations include the detection of polarisation and the presence of two protruding small lobes on opposite sides of some CCSN remnants (termed `Ears') (e.g., \citealt{Wangetal2001, Maundetal2007, Milisavljevic2013, Gonzalezetal2014, Marguttietal2014, Inserraetal2016, Mauerhanetal2017, Bearetal2017, GrichenerSoker2017, Garciaetal2017,  LopezFesen2018}).  The degree to which jets play roles in the explosion mechanism and in the evolution of CCSNe is still an open question. 
 
Neutrinos carry most of the energy that the formation of a neutron star (NS) in core-collapse supernovae (CCSNe) liberates, while the CCSN ejecta carry only a small fraction of that energy. Constructing a theoretical mechanism to convert even a small fraction of the released gravitational energy to kinetic energy of the CCSN ejecta is challenging. 

In one explosion mechanism neutrinos that heat the in-flowing gas revive the stalled shock and explode the star, i.e., the \textit{delayed neutrino mechanism} \citep{BetheWilson1985, Bruennetal2016, Jankaetal2016, Muller2016, Burrowsetal2018, Mabantaetal2019,  Casanovaetal2020, Couchetal2020, DelfanAzarietal2020, Iwakamietal2020, KazeroniAbdikamalov2020, Kurodaetal2020, PowellMuller2020, Stockingeretal2020}. 

In a second  explosion mechanism, jets that the just-born NS (or black hole) launches drive the explosion. Even if the pre-collapse core is slowly rotating (or not at all), the mass that the NS accretes possesses stochastic angular momentum that forms an intermittent accretion disk (or belt) that launches jets in varying directions and intensities. This is the \textit{jittering jets explosion mechanism} (e.g., \citealt{Soker2010, PapishSoker2011, PapishSoker2014, GilkisSoker2014, GilkisSoker2015, Quataertetal2019}). In not requiring pre-collpase core rotation, therefore it might explain most CCSNe, the jittering jets explosion mechanism fundamentally differs from many other cases of jet-driven CCSNe that require pre-collpase rapid core rotation, and therefore are very rare (e.g., \citealt{Khokhlovetal1999, Aloyetal2000, Burrows2007, Nagakuraetal2011, TakiwakiKotake2011, Lazzati2012, Maedaetal2012, Bromberg_jet, Mostaetal2014, LopezCamaraetal2014, BrombergTchekhovskoy2016, LopezCamaraetal2016, Nishimuraetal2017, Fengetal2018, Gilkis2018}). 
In the jittering jets explosion mechanism, unlike in the delayed neutrino mechanism, we expect no failed CCSNe. To the contrary, the collapse of the core to form a black hole will probably lead to a super-energetic CCSN, i.e., an explosion energy of $E_{\rm exp} > 10^{52} \erg$ \citep{Gilkisetal2016Super}. The two explosion mechanisms differ in some of their other  predictions (e.g., \citealt{Gofmanetal2020}). 

Both explosion mechanisms are yet to overcome some challenges (e.g.,\citealt{KaplanSoker2020b}). Interestingly, results from recent years hint that there is some connection between these two mechanisms. Studies of the delayed neutrino mechanism consider pre-explosion perturbations to solve some of the problems of this mechanism (but not all of them, e.g., \citealt{SawadaMaeda2019}), such as in the convective zones of the pre-collapse core  (e.g., \citealt{CouchOtt2013, Mulleretal2019Jittering}).
These lead to stochastic angular momentum of the gas that the NS accretes. Consequently, some three-dimensional (3D) core collapse simulations (e.g., \citealt{Mulleretal2019Jittering}) find outflow morphologies that resemble jittering jets, i.e.,  the axis of the bipolar outflow changes its direction \citep{Soker2019JitSim}. 
From the other side, the jittering jets explosion mechanism seems to require that neutrino heating takes place \citep{Soker2018KeyRoleB, Soker2019SASI}. 

Magnetic fields seem to play important roles in the neutrino driven mechanism(e.g., \citealt{Buglietal2020}), and a critically important role in the jittering jets explosion mechanism (\citealt{Soker2018KeyRoleB, Soker2019SASI}). Magnetic fields are usually not included in simulations, but might further connect the two mechanisms.  

In this study we consider rare cases where the central compact object, a NS or a black hole, launches late jets, weeks to months after explosion, as it accretes fallback material.  These jets are not involved in the explosion itself as they are active after the formation of the central object and after the unbinding of the ejecta. 
We are motivated by recent calculations that suggest that late jets might solve some puzzles in rare CCSNe. 
\cite{KaplanSoker2020a} suggest that jets that the newly-born NS launches weeks to months after explosion might account for peaks in the light curve of some peculiar (i.e. having unusual light curves) CCSNe, such as the peaks in the light curve of iPTF14hls. For their calculations they built a toy model where each of the two jets inflate a bubble (cocoon), but they did not simulate the process of bubble inflation. 
\cite{KaplanSoker2020b} and \cite{SokerKaplan2021} assume a bipolar ejecta morphology, and with a simple modelling calculate the light curve as a result of two opposite low-density  bubbles along the  polar directions, i.e., a bipolar morphology. Again, they did not simulate the bipolar morphology, but rather assumed it. They find that there is a rapid decline in the light curve for an equatorial observer. This might explain the abrupt decline in the light curve of SN~2018don.
 
We conduct three-dimensional (3D) hydrodynamical simulations to explore the process by which the jets inflate bubbles in the ejecta. In section \ref{sec:Numerical} we describe our numerical setting and in section \ref{sec:results} we describe our results. We summarise the main results in section \ref{sec:summary}.

\section{Numerical setup}
\label{sec:Numerical}

We use version 4.2.2 of the adaptive-mesh refinement (AMR) hydrodynamical FLASH code \citep{Fryxell2000} in three dimension (3D). 
As the strong jet-ejecta interaction takes place in optically thick regions, we turn off radiative cooling at any gas temperature.
The equation of state includes both radiation pressure and gas pressure with an adiabatic index of $\gamma=5/3$, due both to ions and electrons, i.e., $P_{\rm tot} = P_{\rm rad} + P_{\rm ion}+P_{\rm elec}$. 
  
We employ a full 3D AMR using a Cartesian grid $(x,y,z)$ with outflow boundary conditions at all boundary surfaces. We use either regular resolution with 
7 refinement levels and a minimum cell size of $\Delta_{\rm cell,m}=2.34 \times 10^{13} \cm$, or high resolution with 8 refinement levels and a minimum cell size of $\Delta_{\rm cell,m}=1.17 \times 10^{13} \cm$. 
We inject the two opposite jets along a constant axis, the $z$-axis. The $z=0$ plane is the equatorial plane of the flow. We simulate the whole space (the two sides of the equatorial plane), with a total size of the Cartesian  numerical grid of $(800 \AU)^3$, i.e., $(L_x,L_y,L_z) = \pm 400 \AU$.

We start with a CCSN ejecta with a mass of $M_{\rm ej}$ and a kinetic energy if $E_{\rm SN}$. 
We take the ejecta a long time after the explosion, such that the initial (when we start the simulation) velocity at each radius is
$v(r)=r/t_0$, where $t_0$ is the time after explosion when we start the simulation. 
We take the initial density profile from \cite{SuzukiMaeda2019} (their equations 1-6, with $l=1$ and $m=10$), which reads 
\begin{equation}
\rho (r, t_0) = \begin{cases}
        \rho_0 \left( \frac{r}{t_0 v_{\rm br}} \right)^{-1} 
        & r\leq t_0 v_{\rm br}
        \\
        \rho_0 \left( \frac{r}{t_0 v_{\rm br}} \right)^{-10} 
        & r>t v_{\rm br}, 
        \end{cases}
\label{eq:density_profile}
\end{equation}
where 
\begin{eqnarray}
\begin{aligned} 
& v_{\rm br} = 1.69 \left( \frac {E_{\rm SN}}{M_{\rm ej}} \right)^{1/2} 
=7.58 \times 10^3 
\\& 
\times \left( \frac {E_{\rm SN}}{2 \times 10^{51} \erg} \right)^{1/2}
\left( \frac {M_{\rm ej}}{5 M_\odot} \right)^{-1/2}
\km \s^{-1} ,
\end{aligned}
\label{eq:vbr}
\end{eqnarray}
and
\begin{equation}
\rho_0 = \frac {7 M_{\rm ej}}{18 \pi v^3_{\rm br} t^3_0} .
\label{eq:rho0}
\end{equation}
To avoid numerical difficulties near the center we set an inner sphere at $r<R_{\rm in}= 10^{14} \cm$ to have a constant density. Namely, $\rho(r<R_{\rm in}) = \rho(R_{\rm in})$. 
In all cases that we simulate the explosion energy is $E_{\rm SN} = 2 \times 10^{51} \erg$ and the ejecta mass is $M_{\rm ej} = 5 M_{\odot}$.

 In most simulations  we launch the two jets in two opposite cones from the inner $\Delta r_{\rm j}=3 \times 10^{14} \cm $ zone along the $z$-axis ($x=y=0$) and  within a half opening angle of $\alpha_{\rm j} = 20^\circ$  (in one simulation we have $\alpha_{\rm j} =10^\circ$ and in one simulation we have $\alpha_{\rm j} = 50^\circ$).  Although we expect jets' velocity of $v_{\rm j} \ga 10^5 \km \s^{-1}$, to save computational resources we take $v_{\rm j} = 5 \times 10^4 \km \s^{-1}$. 
At the beginning of the simulation the two opposite cones are filled with the jets material. This implies that the jets are already active for a time period of $\Delta t_{\rm j,0}=\Delta r_{\rm j}/v_{\rm j}=6 \times 10^4 \s = 0.694 \days$.  
We continue to inject the jets for a time period of $\Delta t_{\rm j,a}$, such that the jets are active for a total time period of $\Delta t_{\rm j}=\Delta t_{\rm j,0} + \Delta t_{\rm j,a}$.
 To test the effects of the value of $\Delta r_{\rm j}$, 
 in one high-resolution simulation we take $\Delta r_{\rm j}=1.5 \times 10^{14} \cm $, namely, half the fiducial value.  
 
In simulations HR, HR1/2, RR10, RR20, RR50, and HE we take $\Delta t_{\rm j,a}=0.1 \days$. Overall, the jets are active for $\Delta t_{\rm j}=0.79 \days$ in these two cases. The mass loss rates into the two jets in the first 5 simulations is $\dot M_{\rm 2j}= 8 \times 10^{-4} M_\odot ~{\rm day}^{-1}$, and it is $\dot M_{\rm 2j}= 0.02 M_\odot ~{\rm day}^{-1}$ in the HE simulation; 
the total energies in the two jets are $E_{\rm 2j} = 1.6 \times 10^{49} \erg=0.008 E_{\rm SN}$ and $E_{\rm 2j} = 4 \times 10^{50} \erg=0.2 E_{\rm SN}$, respectively. 

We also run two long-activity cases where the duration of the jet activity is $\Delta t_{\rm j}=50 \days$, with a total energy of $E_{\rm 2j} = 1.6 \times 10^{49} \erg=0.008 E_{\rm SN}$ (LA) and 
$E_{\rm 2j} = 4 \times 10^{50} \erg=0.2 E_{\rm SN}$ (LAHE). 
We summarise the simulations we perform in Table \ref{Table:cases}. 
\begin{table}
\centering
\begin{tabular}{|c|c|c|c|c|c|}
\hline
 Case  & $\Delta_{\rm cell,m}$ & $E_{\rm 2j}$ & $\Delta t_{\rm j}$ & $\alpha_{\rm j}$ &Figures \\ 
   & cm  & erg & days   & & \\ 
 \hline 
HR & $1.17 \times 10^{13}$ & $1.6 \times 10^{49}$ & $0.79$ & $20^\circ$ & \ref{fig:DensityMarks} - \ref{fig:HRCase1Evolution}\\ \hline
HR1/2 & $1.17 \times 10^{13}$ & $1.6 \times 10^{49}$ & $0.79$ & $20^\circ$ & \ref{fig:HRhalf}\\ \hline
RR10 & $2.34 \times 10^{13}$ & $1.6 \times 10^{49}$ & $0.79$ & $10^\circ$  & \ref{fig:Angles} \\ \hline
RR20 & $2.34 \times 10^{13}$ & $1.6 \times 10^{49}$ & $0.79$ & $20^\circ$  & \ref{fig:Angles}, \ref{fig:NonSelf} \\ \hline
RR50 & $2.34 \times 10^{13}$ & $1.6 \times 10^{49}$ & $0.79$ & $50^\circ$  & \ref{fig:Angles} \\ \hline
HE &$2.34 \times 10^{13}$  & $4 \times 10^{50}$   & $0.79$ & $20^\circ$  & \ref{fig:LR25energy}\\ \hline
LA & $2.34 \times 10^{13}$& $1.6 \times 10^{49}$   & $50$ & $20^\circ$  & \ref{fig:LA50days}\\ \hline
LAHE &$2.34 \times 10^{13}$  & $4 \times 10^{50}$   & $50$ & $20^\circ$  & \ref{fig:LAHE50days} -  \ref{fig:NonSelf} \\ \hline
\end{tabular}
\caption{Summary of the distinguish properties of the high-resolution simulations (HR and HR1/2), and the regular resolution simulations,  including varying $\alpha_{\rm j}$ values (RR),  high-energy (HE), long-activity (LA), and long-activity high-energy (LAHE). In the second column we list the minimum cell size in the numerical grid. The third column gives the total energy in the two jets and the fourth column gives the time period of jets' activity, including $\Delta t_{\rm j,0}=0.694 \days$ before we start the simulation,  beside simulation HR1/2 for which $\Delta t_{\rm j,0}=0.347 \days$. 
 In the fifth column we list the values of $\alpha_{\rm j}$, the half opening angle of the jets. 
In all simulations we inject the jets  with a velocity of $v_{\rm j} = 5 \times 10^4 \km \s^{-1}$, the explosion energy (kinetic energy of the ejecta) is $E_{\rm SN} = 2 \times 10^{51} \erg$, and the ejecta mass is $M_{\rm ej} = 5 M_{\odot}$.   }
\label{Table:cases}
\end{table}

For numerical reasons (to avoid very low densities) we inject a very weak slow ($500 \km \s^{-1}$) wind in the directions where we do not launch the jets, i.e., in the sector $\alpha_{\rm j} < \theta \le 90^\circ$ in each hemisphere.  This flow carries only $10^{-5}$ times the mass in the jets, so both negligible amounts of mass and energy.   Because of the constant-density sphere near the center, and the region where we numerically inject the jets, the flow structure close to the center includes some numerical effects.
The initial temperature of the simulation box and the jets is $10000\K$.

We find the radius of the photosphere of the ejecta itself from the relation 
\begin{equation}
\tau = \int_{r_i}^\infty \kappa \rho dr = \frac{2}{3}.
\label{eq:tau}
\end{equation}
We first check at each relevant time whether the photosphere is at $r>v_{\rm br} t$. In this case the photosphere is at $r_{\rm ph}=r_1>v_{\rm br} t$, where  
\begin{eqnarray}
\begin{aligned} 
& r_1 = 9.7 \times 10^{14} 
    \left(\frac{\kappa}{0.3 \cm^2 \g^{-1}} \right)^{1/9} 
    \left(\frac{E_{\rm SN}}{10^{51} \erg}\right)^{7/18}
   \\&  \times 
    \left(\frac{M_{\rm ej}}{5 M_{\odot}}\right)^{-5/18}
    \left( \frac{t}{10^6 \s} \right)^{7/9} \cm;  \qquad {\rm for} \quad \tau (v_{\rm br} t) > 2/3 . 
\end{aligned}
\label{eq:r1}
\end{eqnarray}
If the outer part of the ejecta is optically thin, i.e., $\tau (v_{\rm br} t) < 2/3$, we neglect the contribution of the outer part (gas at $r>v_{\rm br} t$), and consider only the contribution of the inner part of the power law to the optical depth. This gives the photosphere at $r_i =r_2$, where  
\begin{eqnarray}
\begin{aligned} 
r_2 =  & 5.36  \times 10^{14}  
    \left(\frac{E_{\rm SN}}{10^{51} \erg}\right)^{1/2}  
    \left(\frac{M_{\rm ej}}{5 M_{\odot}}\right)^{-1/2} \left( \frac{t}{10^6 \s} \right)
    \\ \times & \exp 
    \left[ - 5.19 \times 10^{-4}   
    \left(\frac{\kappa}{0.3 \cm^2 \g^{-1}}\right)^{-1} \right.
    \\ & \times \left.
    \left(\frac{E_{\rm SN}}{10^{51} \erg}\right) 
    \left(\frac{M_{\rm ej}}{5 M_{\odot}}\right)^{-2} \left( \frac{t}{10^6 \s} \right)^2 \right] \cm.
\end{aligned}
\label{eq:r2}
\end{eqnarray}

\section{Results}
\label{sec:results}

We have two goals. Firstly, we want to check the general morphology and characteristics of the jet-ejecta interaction (section \ref{subsec:Flow}), and secondly, to determine the location of the jet-inflated bubbles with respect to the (approximate) photosphere as function of time (sections \ref{subsec:Evolution} and \ref{subsec:OtherCases}). 

\subsection{The basic flow structure: The HR simulation}
\label{subsec:Flow}

We first describe a high-resolution (HR) simulation that we summarise in table \ref{Table:cases}. Because high resolution simulations demand large computer resources, we have only two simulations of high resolution. We present some of the results of simulation HR below, and postpone the discussion of other aspects and the presentation of other simulations to later subsections. 
In what follows we measure the time from the explosion (beside in Fig. \ref{fig:Pevolution}). 

\begin{figure}
	\centering
\includegraphics[trim=3.6cm 8.0cm 2.5cm 4.5cm ,clip, scale=0.55]{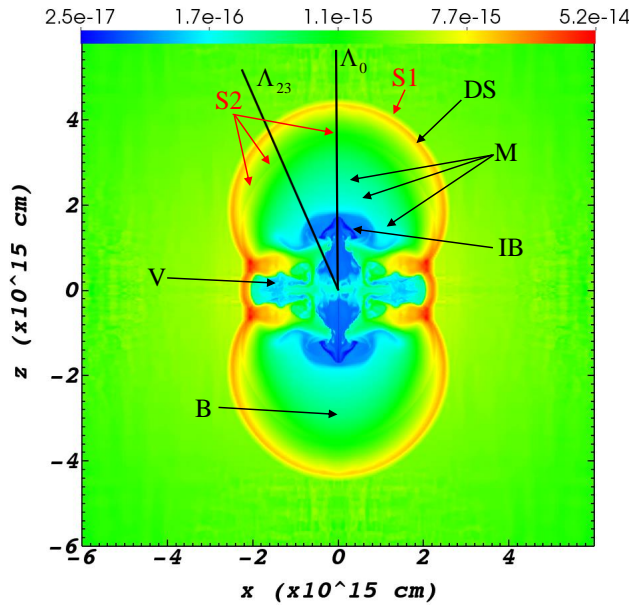}
	\caption{A density map in the meridional plane of the high resolution (HR) run at $t=154 \days$. At that time the jets are not active anymore. The $z$-axis is the symmetry axis of the jets. The color-bar gives the density in units of $\g \cm^{-3}$. We mark (only on one side of the equatorial plane $z=0$) the strong forward shock (S1), a weak inner shock (S2), a region where jets' material mixes with the ejecta material (M), large vortexes (V), the dense shell (DS), the bubble (B) that is the volume inner to the dense shell, the inner bubble (IB) that is the shocked jets' material, and the two lines along which we present physical quantities, in Fig. \ref{fig:Line0} for line $\Lambda_0$ along the symmetry axis, and in Fig. \ref{fig:Line23} for the line $\Lambda_{23}$ along a direction of $23^\circ$ to the symmetry axis. 
	}
	\label{fig:DensityMarks}
\end{figure}

\begin{figure}
	\centering
\includegraphics[trim=1.7cm 5.9cm 0.0cm 3.5cm ,clip, scale=0.42]{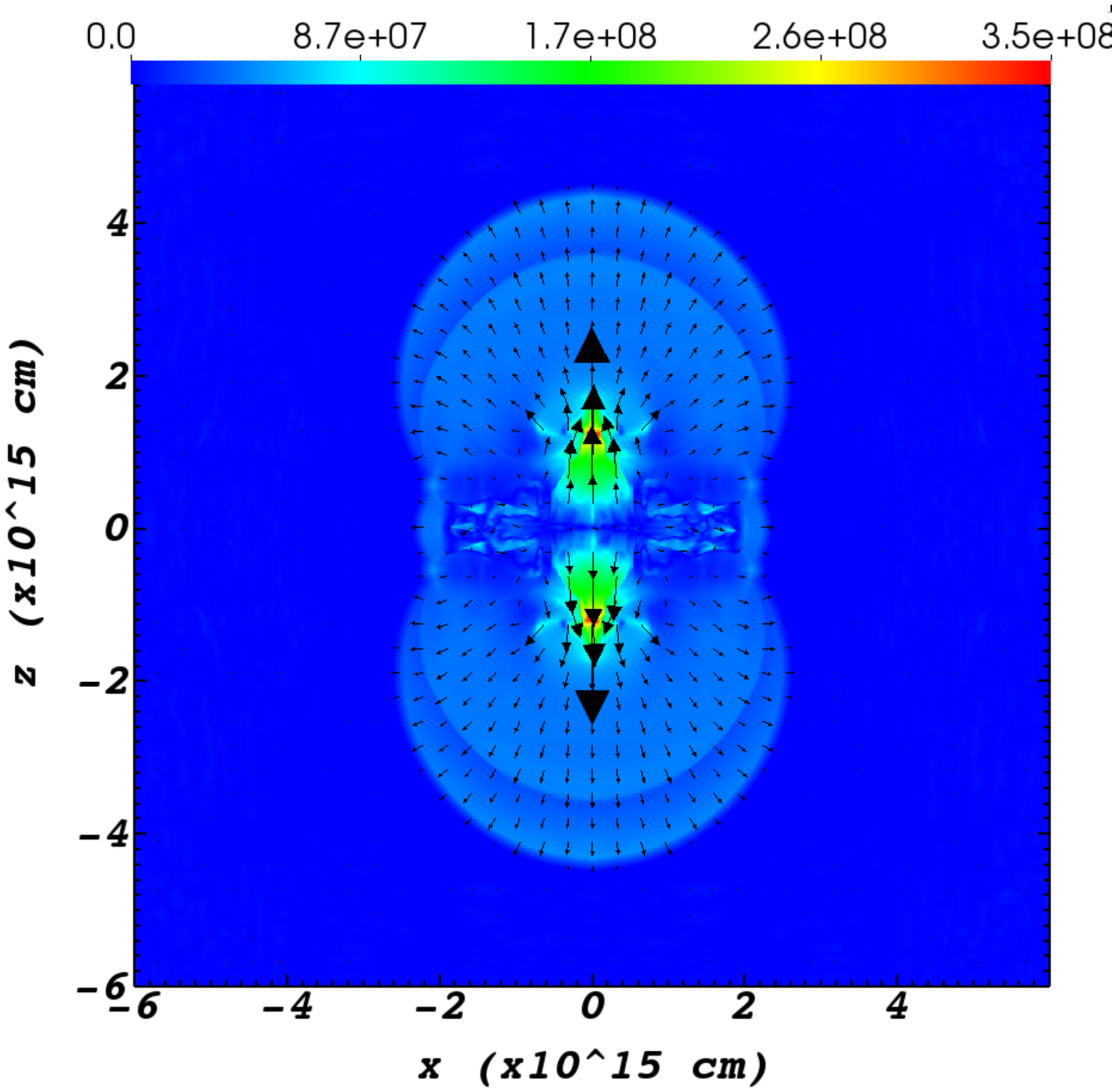}
	\caption{The same as in Fig. \ref{fig:DensityMarks} but showing the velocity relative to the homologous expansion, $v_{\rm rel}$ from equation (\ref{eq:Vrel}). Colours indicate the magnitude of the velocity according to the colour-bar in $\cm \s^{-1}$.  
	}
	\label{fig:VelocityMapHR}
\end{figure}

\begin{figure}
	\centering
\includegraphics[trim=0.2cm 6.0cm 0cm 7.0cm ,clip, scale=0.42]{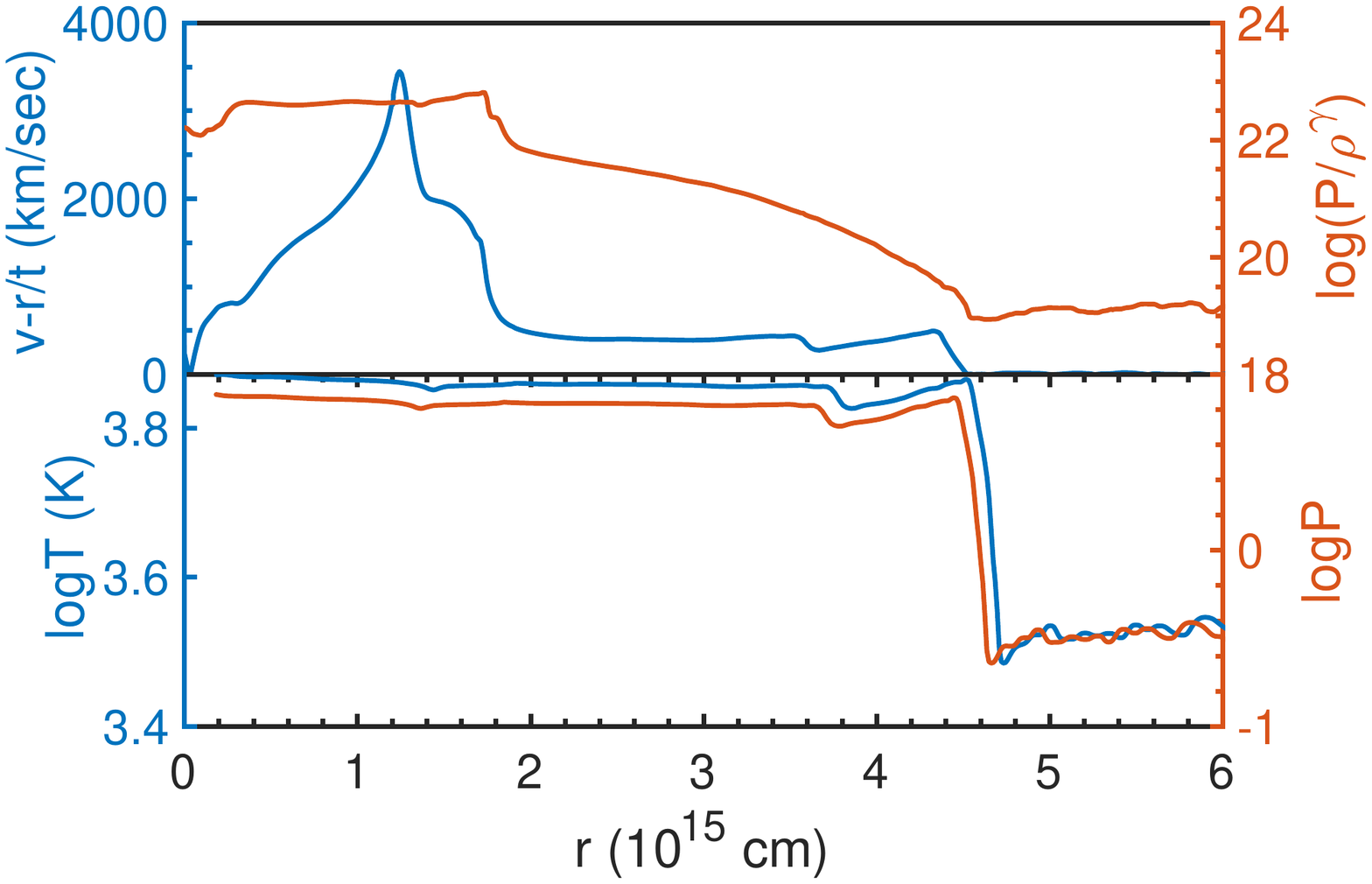}
\caption{Several quantities as function of distance from the center and along the symmetry axis, the line $\Lambda_0$ that we mark on Fig. \ref{fig:DensityMarks}. Upper panel: The radial velocity relative to the homologous velocity $v_{\rm rel,r}$ from equation (\ref{eq:Vrel}) and the function $\log P/\rho^{\gamma}$ in cgs units that represents entropy. Lower panel: Temperature and pressure in log scale and in cgs units. 
We identify shocks at $r\simeq  4.5 \times 10^{15} \cm$ and at $r\simeq  3.6 \times 10^{15} \cm$.
	}
	\label{fig:Line0}
\end{figure}

\begin{figure} 
	\centering
\includegraphics[trim=0cm 6.0cm 0cm 5.0cm ,clip, scale=0.42]{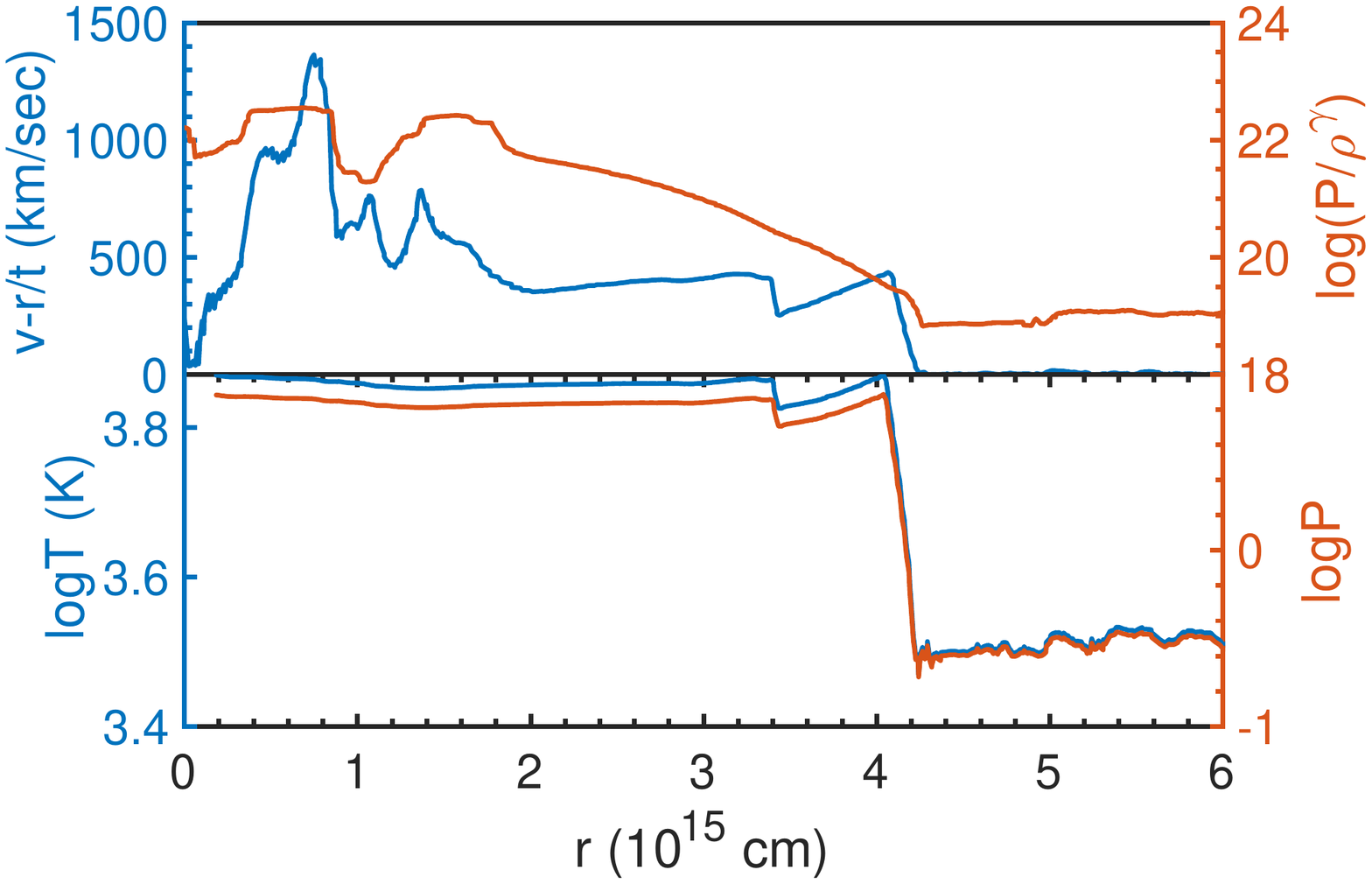}
\caption{Similar to Fig. \ref{fig:Line0} but for the line $\Lambda_{23}$ that is at $23^\circ$ from the symmetry axis as we mark on Fig. \ref{fig:DensityMarks}.  We identify shocks at $r\simeq  4.2 \times 10^{15} \cm$ and at $r\simeq  3.4 \times 10^{15} \cm$
	}
	\label{fig:Line23}
\end{figure}

\begin{figure*} 
	\centering
\includegraphics[trim=4.1cm 12.8cm 1.0cm 2.4cm ,clip, scale=1.4]{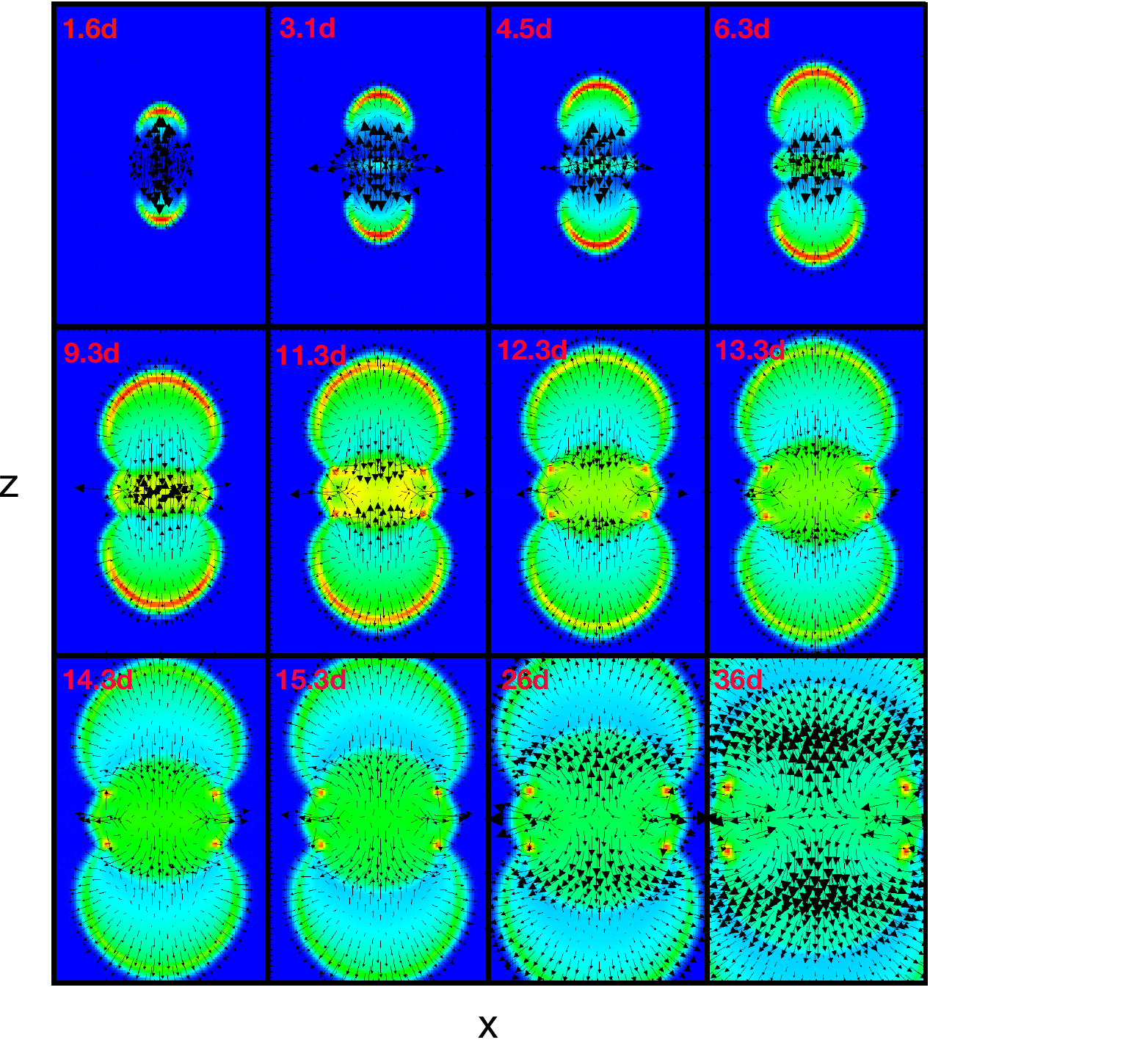}
\caption{Pressure maps with relative velocity (equation \ref{eq:Vrel}) arrows at twelve early times after the beginning of the simulation, as we indicate in days. Note that the times we list in the panels are from the beginning of the simulation that occurs at $t_0=50 \days$. Namely, the twelve frames cover the time period $t=51.6 \days$ to $t=86 \days$. The blue, pale blue, green, yellow, and red colours depict the pressure from lowest value to highest value, respectively.  As three examples, the maximum (minimum) pressures at $t=3.1$, $t=11.3$ and $t=15.3 \days$ are (in $\erg \cm^{-3}$) $P_{\rm max} = 6.2 \times 10^4$ ($P_{\rm min}= 1.25 $), $P_{\rm max} = 2470$ ($P_{\rm min} = 2 $), and 
$P_{\rm max} = 1300 $ ($P_{\rm min}= 2.4 $), 
(for more on the typical values of pressure see Figs. \ref{fig:Line0} and \ref{fig:Line23}). The arrows depict the relative velocity, with their length proportional to the velocity. The typical value of the maximum relative back-flow velocities is $\approx 4 \times 10^4 \km \s^{-1}$.  Each panel extends from $x=-10^{15} \cm$ to $x=10^{15} \cm$ and 
from $z=-1.5 \times 10^{15} \cm$ to $z=1.5 \times 10^{15} \cm$.
	}
	\label{fig:Pevolution}
\end{figure*}

In Fig. \ref{fig:DensityMarks} we present the density in the meridional plane. We take the $z$-axis along the symmetry axis of the jets. In Fig. \ref{fig:VelocityMapHR} we present the map of the velocity relative to the homologous velocity , namely 
\begin{equation}
\vec {v}_{\rm rel} = \vec {v} - \frac{r}{t} \hat{r}. 
\label{eq:Vrel}
\end{equation}
In Figs \ref{fig:Line0} and \ref{fig:Line23} we present the variation of four quantities with the distance from the center and along the two lines $\Lambda_0$ and $\Lambda_{23}$, respectively, as we mark on Fig. \ref{fig:DensityMarks}. Figures \ref{fig:DensityMarks}-\ref{fig:Line23} are all at the same time $t= 154 \days$.

From Figs. \ref{fig:DensityMarks}-\ref{fig:Line23} we identify the following features. We clearly see two low-density regions (in deep blue), one at each side of the equatorial plane, that we refer to as inner-bubbles (IB). These are the post-shock jets' material. In these low-density inner-bubbles the fraction of material that originated from the jets in each numerical cell is $\ga 50 \%$; the rest is ejecta gas that mixed with it. Mixing of jet and ejecta gases takes place in the regions we mark by `M' (pale blue), where the fraction of original jets' material is between few percent to about 50 per cent; the rest is the original ejecta gas. 

We identify backward flow near the equatorial plane (relative to the homologous expansion; Fig. \ref{fig:VelocityMapHR}). These form vortexes; we mark this zone with `V' on Fig. \ref{fig:DensityMarks}. As we indicated in section \ref{sec:Numerical}, we do not fully trust the numerical results as far as quantitative values near the center are concerned, but we do trust the qualitative flow structure. Namely, we do think that the jets-ejecta interaction forms vortexes, but we do not trust their exact quantitative structure. 
 
We can notice two shocks in the figures, the forward strong shock (S1 on Fig. \ref{fig:DensityMarks}) and a weaker shock that trails the forward shock (S2). These shock fronts are clearly visible in Fig. \ref{fig:VelocityMapHR} as two sharp outward velocity drops, and in Figs. \ref{fig:Line0} and \ref{fig:Line23} as pressure, temperature, and velocity sharp changes.  We discuss the formation of the shock S2  later in this section. 
Behind the forward shock there is a thin dense shell (DS). The entire volume inner to the dense shell is the bubble (B), one bubble at each side of the equatorial plane.
We note that the interaction suffers from Rayleigh–Taylor instability modes during the activity phase of the jets. There are instabilities in the boundary of the dense shell as in this region the pressure gradient and the density gradient have opposite signs. However, even those instability features that grew during the jet activity phase are smoothed by the flow inside the bubble. Later we will see that when the jets are long-lived, instabilities develop around the dense shell.   
 
\begin{figure} 
	\centering
\includegraphics[trim=0cm 7.0cm 0cm 7.0cm ,clip, scale=0.42]{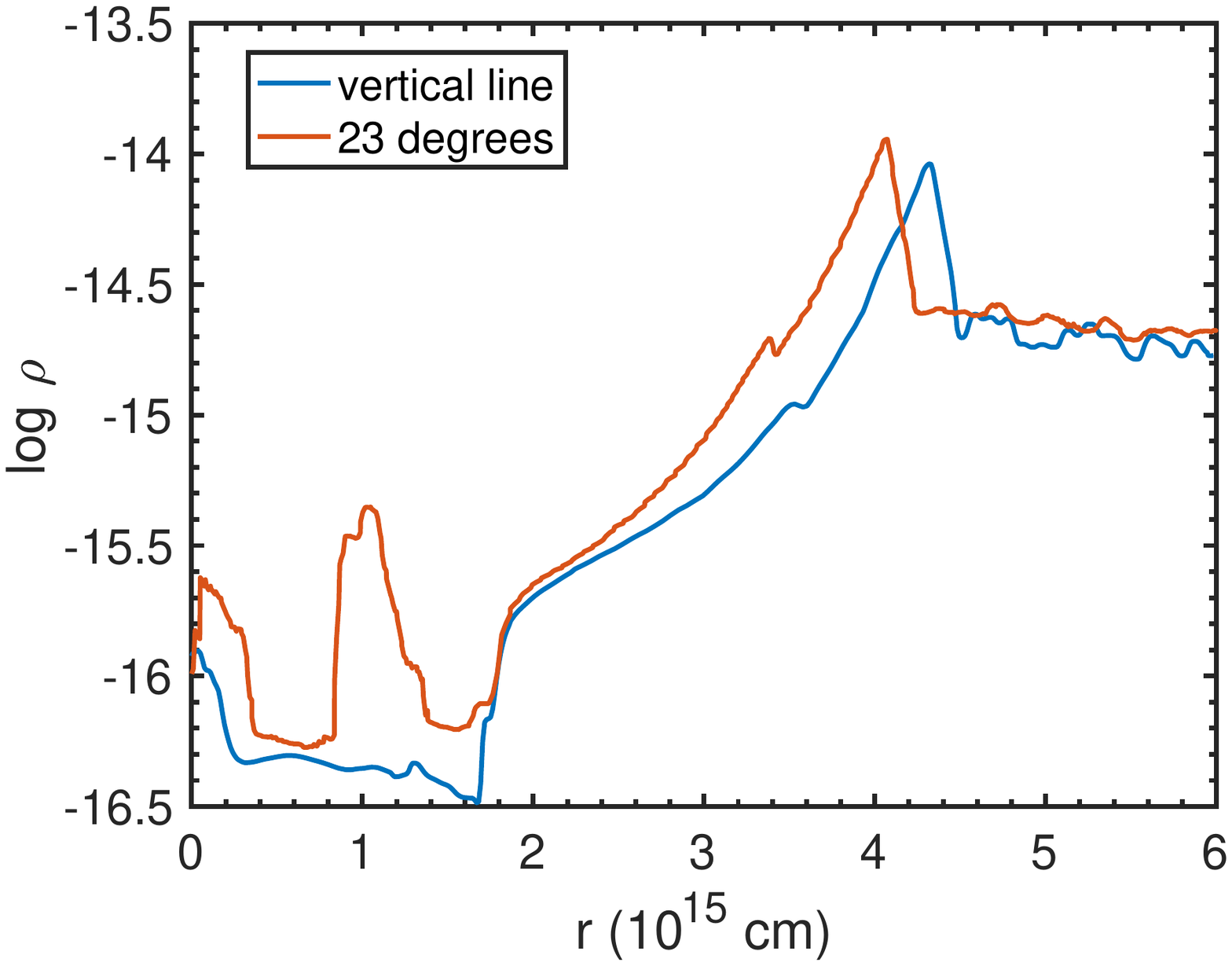}
\caption{The density (in log scale and in units of $\g \cm^{-3}$) profiles along the radial lines $\Lambda_{0}$ (vertical) and $\Lambda_{23}$, as we mark on Fig. \ref{fig:DensityMarks}.
	}
	\label{fig:LinesDensity}
\end{figure}

To elaborate on the density structure we present in Fig. \ref{fig:LinesDensity} the density profiles (in log scale and in units of $\g \cm^{-3}$) along the lines $\Lambda_0$ and $\Lambda_{23}$. Together with Fig. \ref{fig:DensityMarks} we identify the following density structures. The outer most region at $r \ga 4.5 \times 10^{15} \cm$ along the $x=y=0$ line ($\Lambda_0$) and at closer distances in other directions, is the homologous expanding ejecta that the jets did not influence the flow of yet.  The wiggles in the density profile at $r \ga 4.5 \times 10^{15} \cm$ in Fig. \ref{fig:LinesDensity} are due to numerical effects that we detect also in a simulation without jets.  A dense and a relatively thin shell (DS in Fig. \ref{fig:DensityMarks}) behind the forward shock (S1) bounds from inside the undisturbed ejecta. The dense shells has a bipolar morphology, i.e., two opposite bubbles. Behind the dense shell the density drops to below its value had there were no jets. This has an implication for the light curve as we discuss later \citep{KaplanSoker2020b, SokerKaplan2021}. Near the center, within $r \la 1.8 \times10^{15} \cm$, we find the two inner-bubbles, i.e., the very-low-density zones ($\rho < 10^{-16} \g \cm^{-1}$; coloured blue in Fig. \ref{fig:DensityMarks}), one at each side of the equatorial plane. Each inner-bubble  has the shape of a mushroom.  

The weak S2 shock that trails the forward shock (Fig. \ref{fig:DensityMarks}) results from an early-time fallback flow that creates a high-pressure region in the center. To follow that evolution, we present in Fig. \ref{fig:Pevolution} twelve frames of pressure and velocity maps. Only in this figure the time in each frame is the time from the beginning of the simulation $t_0$, rather than the time from explosion.  

The evolution proceeds as follows. The jets empty the center, and for the first day or so the pressure in the center is very low, while regions of high pressure are developing behind the forward shock (the 1.6d frame in Fig. \ref{fig:Pevolution}). By about few days after jets' injection the high pressure pushes gas back toward the center. We see a back-flow in the frame 3.1d. This back-flow collides with itself near the center and it forms a large high-pressure region. After several days there is an outflow in and near the equatorial plane. In the frame 9.3d we see the full inflow stream, with a maximum back-flow velocity of $v_{\rm back} \simeq 4 \times 10^4 \km \s^{-1}$. In the frame at 11.3d the pressure in the center reaches its maximum value (yellow color). From frame 12.3d on, the back-flow decreases, and an outflow in the polar directions develops. In the last frame, 36d, we see a full polar outflow, that later forms the trailing shock S2. We also notice the development of the vortexes near the equatorial plane. 
  
\subsection{Evolution and implications on light curves}
\label{subsec:Evolution}

We neither calculate the effects on the light curve nor we include radiative transfer. Such calculations will have to include, in addition to radiative transfer, recombination of the ejecta, calculation of the opacity at each point, and radioactive nuclei. As well, we assume a constant opacity. For that, the location of the photosphere is a very crude estimate. Nonetheless, it serves our purpose of presenting the general behavior of late jets that interact with the ejecta. The study of the influence of the jets on the light curve is a topic of a follow-up paper. 

\begin{figure} 
\centering
\begin{tabular}{cc}
\includegraphics[trim=0.0cm 6.0cm 0cm 5.2cm ,clip, scale=0.34]{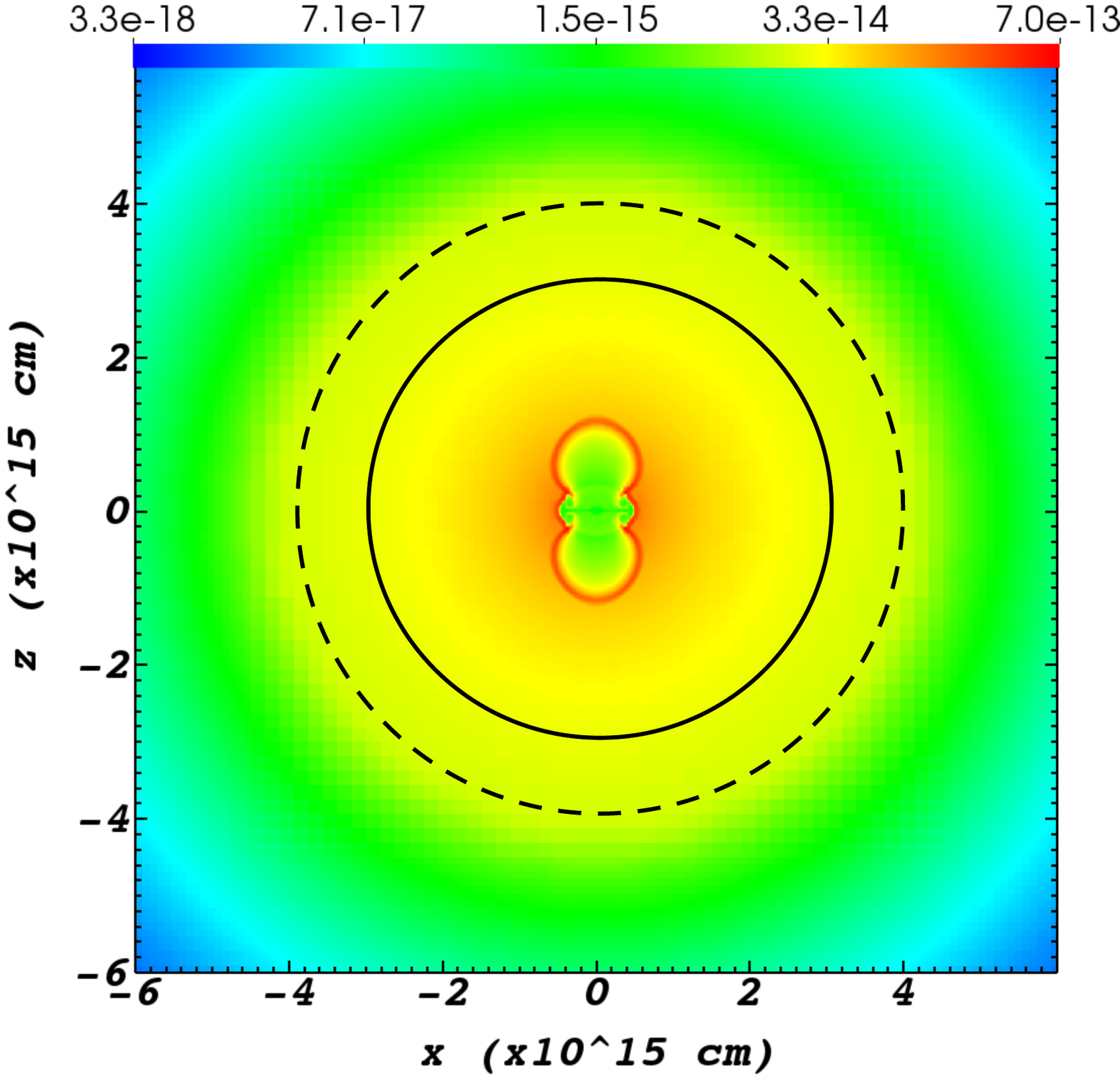} \\
\includegraphics[trim=0.0cm 6.0cm 0cm 5.2cm ,clip, scale=0.34]{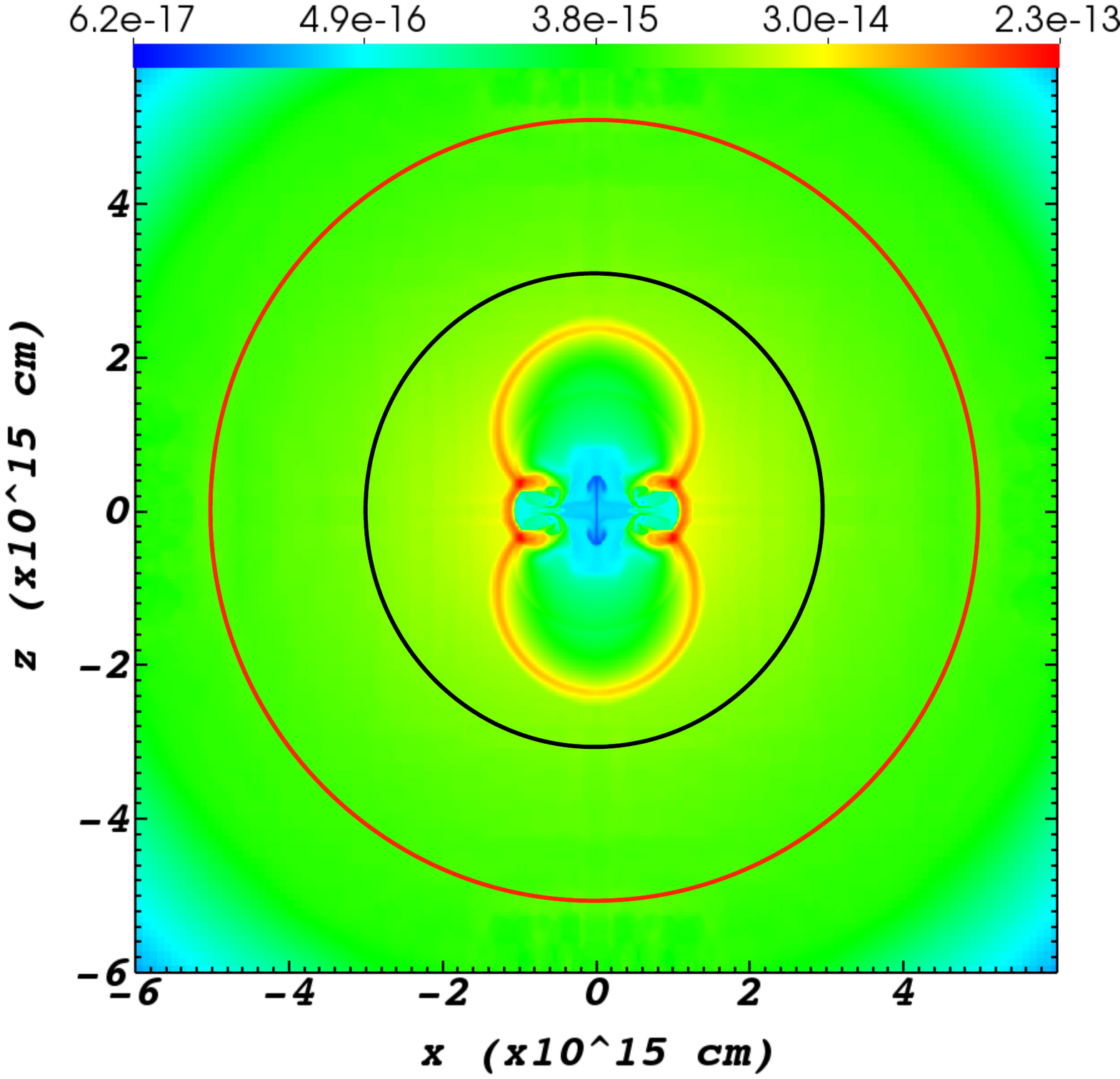}\\
\includegraphics[trim=0.0cm 6.0cm 0cm 5.2cm ,clip, scale=0.34]{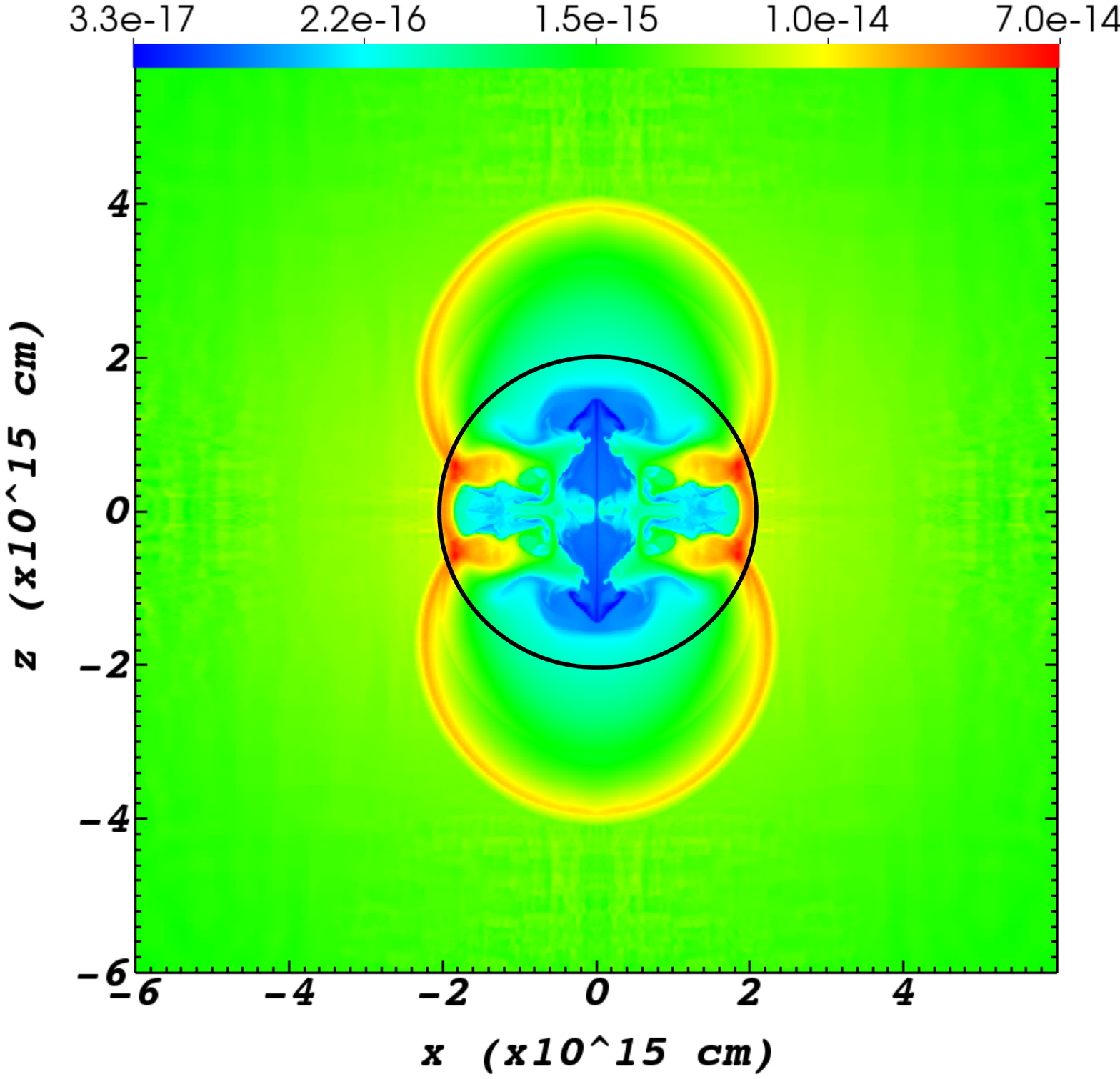} 
\end{tabular}
\caption{ Density maps in the meridional plane of the HR simulation at three times, from top to bottom, $t=62 \days$,  $t=96 \days$, and $t=142 \days$ after explosion. The density scale is according to the color bar in units of $\g \cm^{-3}$. The solid-black circle in each panel marks the photosphere according to our crude estimate by equation (\ref{eq:r2}) and for an opacity of $\kappa=0.03$, while the solid-red circle in the middle panel is for $\kappa=0.1$. The dashed circle in the upper panel is the radius where the power-law density profile changes, i.e., $v_{\rm br} t$ (equations \ref{eq:density_profile} and \ref{eq:vbr}; at later times it is outside the numerical grid). 
}
  \label{fig:HRCase1Evolution}
    \end{figure}

In Fig. \ref{fig:HRCase1Evolution} we present density maps at three times for the HR simulation.
We calculate the location of the photosphere by equation (\ref{eq:r2}) and for an opacity of $\kappa=0.03$ in the three panels (solid-black circles) and for $\kappa=0.1$ in the middle panel (solid-red circle). We ignore the presence of the bipolar structure, and so when the photosphere is inside the bubbles the calculation is very crude. (Calculating the exact location of the photosphere in the bubbles requires the inclusion of recombination and the calculation of the opacity in this complicated geometry.)  The dashed-black circle is the location of the break in the power-law density profile that is at $r= v_{\rm br} t$  (equation \ref{eq:density_profile}). At late times the break is outside the numerical grid. 
  
Despite the crude calculation of the location of the photosphere, the evolution in Fig. \ref{fig:HRCase1Evolution} presents important features. These features result from that the photosphere first reaches, from outside, the polar regions of the bipolar structure.
\cite{KaplanSoker2020a} discussed these features. They built toy models to estimate some effects on the light curve, but did not calculate the morphology of the jets-ejecta interaction. Our simulations show the geometrical evolution that they assumed, and allow us to present these features in a clearer way. 

\textit{(1) Energising a peak in the light curve}. We learn from Figs. \ref{fig:Line0} and \ref{fig:Line23} that the temperatures of the dense shells and the bubbles are higher than those of the undisturbed ejecta. Photons from these hotter regions can diffuse out before even the photosphere recedes to the dense shells. These photons might lead to a peak in the light curve of the CCSN \citep{KaplanSoker2020a}. When later the photosphere recedes into the bubbles that are hotter and less dense than the ejecta, the emission might lead to a blue peak (i.e., the extra energy is at shorter wavelength than the rest of the ejecta;  
\citealt{KaplanSoker2020b}).
  
\textit{(2) Rapid light-curve drop for an equatorial observer}.
An observer in and near the equatorial plane ($z=0$) might observe a rapid luminosity decline in the light curve \citep{KaplanSoker2020b, SokerKaplan2021} as a result of a faster recession of the photosphere inside the bubbles. At later times an observer near the equatorial plane will not see the photosphere in the  polar directions. This reduces the flux the observer measures relative to a spherical explosion.   \cite{KaplanSoker2020b} assume that the density in the bubbles is lower than in the ejecta. We here show this. 

\begin{figure} 
\centering
\begin{tabular}{cc}
\includegraphics[trim=0.0cm 6.0cm 0cm 5.2cm ,clip, scale=0.34]{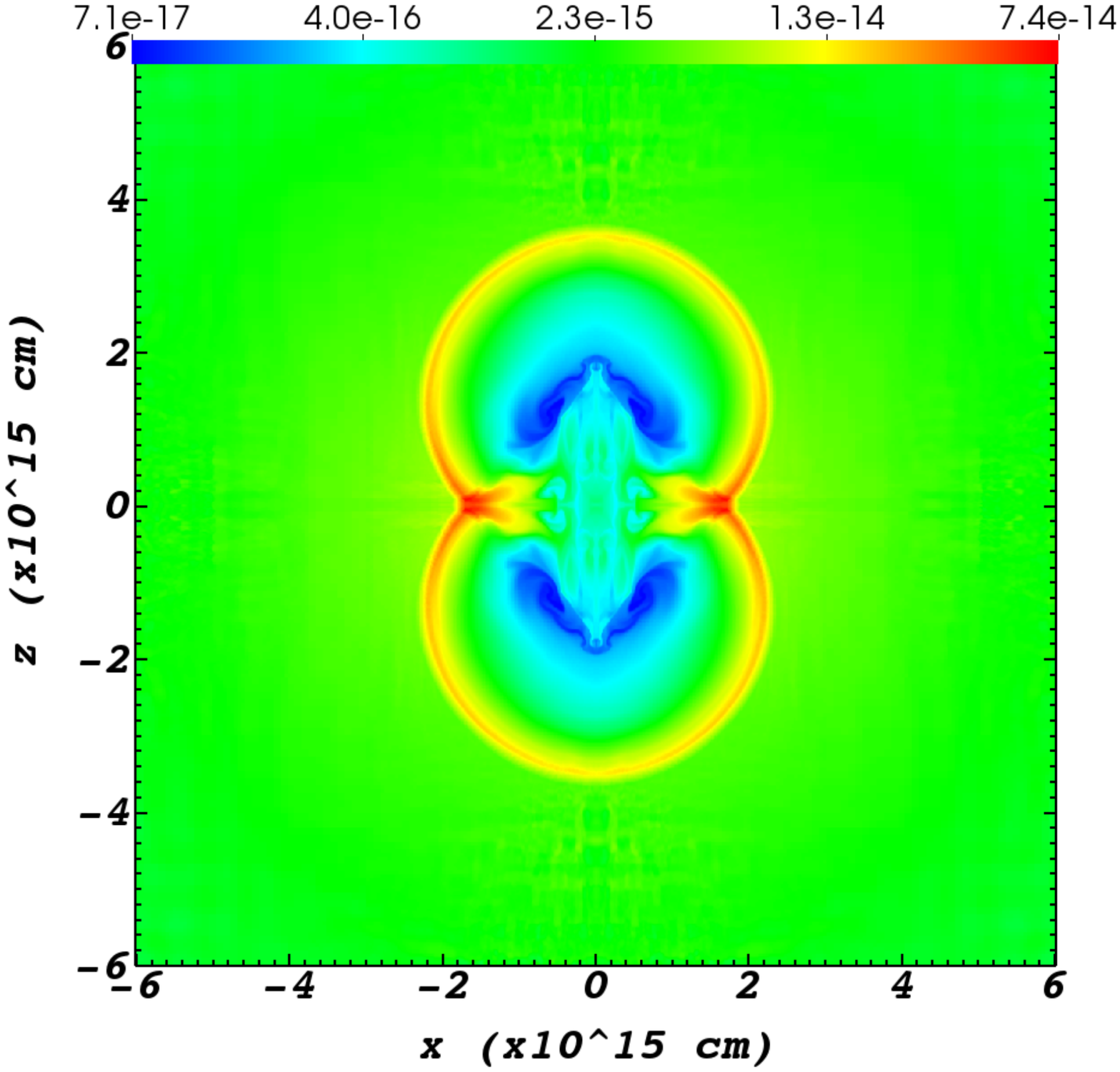} \\
\end{tabular}
\caption{  Density maps in the meridional plane of simulations HR1/2 where the injection length of the jets is half that in simulation HR, but otherwise the two simulations are identical.  Time is $t=142 \days$, so this map can be compared to the bottom panel of Fig. \ref{fig:HRCase1Evolution}.
}
  \label{fig:HRhalf}
    \end{figure}

 We end the study of the high resolution simulations by conducting a simulation where the length of the jets' injection zone is $\Delta r_{\rm j}=1.5 \times 10^{14} \cm $, namely, half the fiducial value. We present the results of this simulation HR1/2 in Fig. \ref{fig:HRhalf}. We cannot take a shorter injection length as we will face numerical difficulties near the center. Comparing Fig. \ref{fig:HRhalf} of simulation HR1/2 with the lower panel of Fig. \ref{fig:HRCase1Evolution} of simulation HR we note the following. First, we see that the general bipolar structure is similar. The structure near the equatorial plane is somewhat different. But, as said, the flow close to the equatorial plane suffers from numerical effects. The size of bipolar structure in simulation HR is larger by about $12\%$. It might be that for numerical reasons we practically inject less energy. We conclude that the injection length of the jets has some influence on the size of the lobes, but not a critical one.   

The main result from the one case we analysed in sections \ref{subsec:Flow} and \ref{subsec:Evolution} is a support to the toy models and conclusions of \cite{KaplanSoker2020a} and \cite{KaplanSoker2020b}. We turn to examine other cases with jets.

\subsection{Varying the opening angle of the jets}
\label{subsec:Angles}

\begin{figure} 
\centering
\begin{tabular}{cc}
\includegraphics[trim=0.0cm 6.0cm 0cm 5.2cm ,clip, scale=0.34]{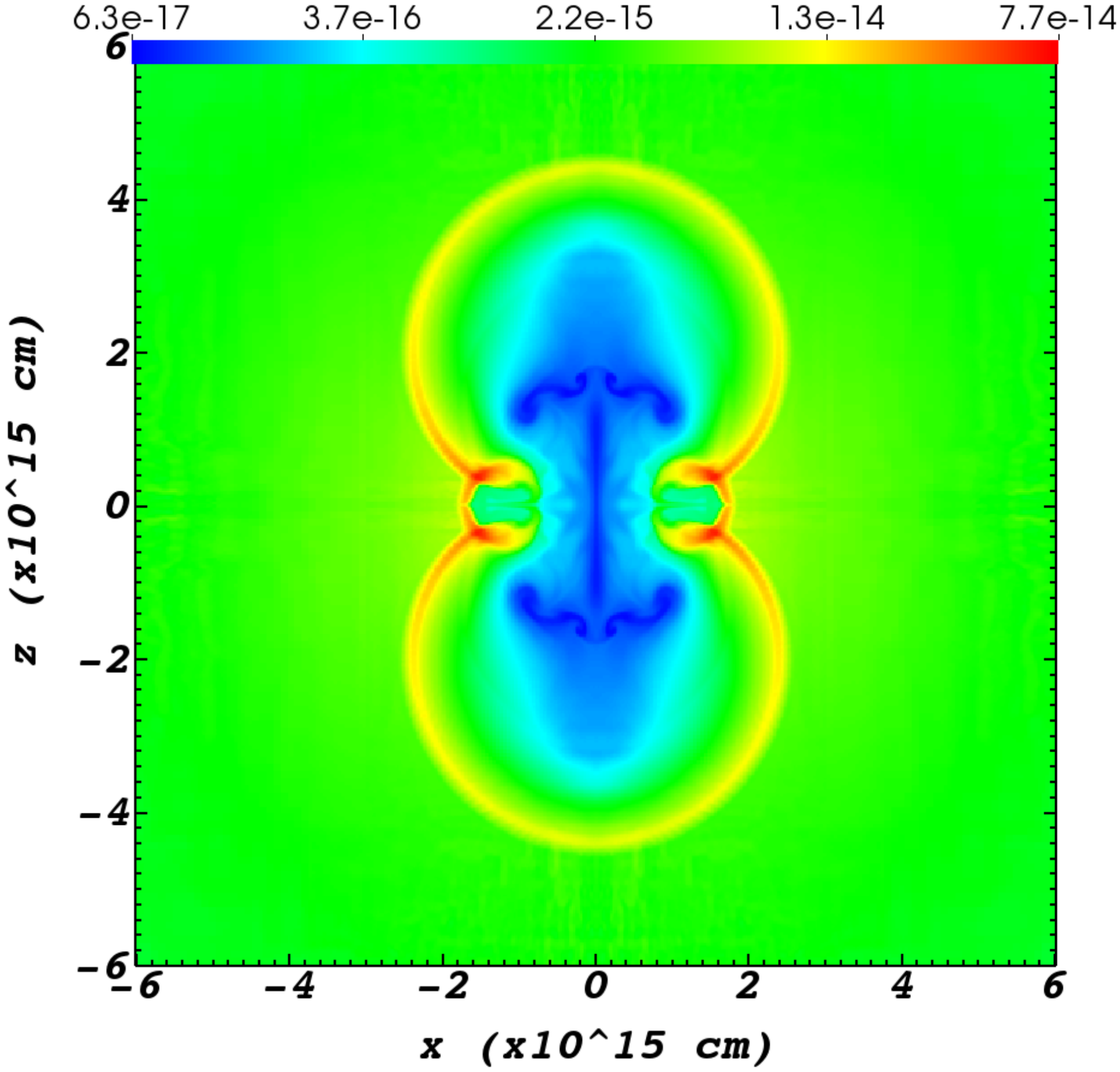} \\
\includegraphics[trim=0.0cm 6.0cm 0cm 5.2cm ,clip, scale=0.34]{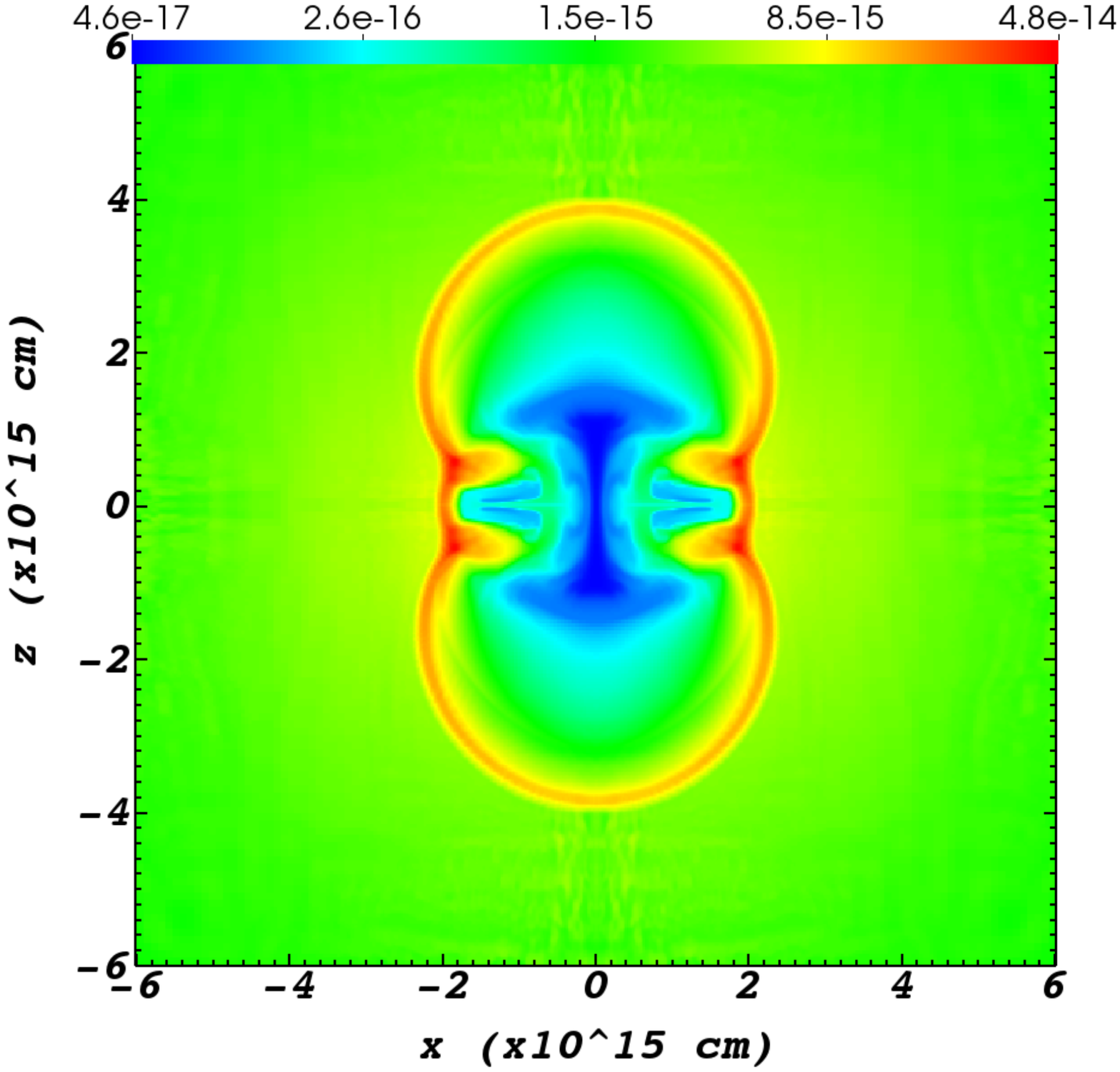}\\
\includegraphics[trim=0.0cm 6.0cm 0cm 5.2cm ,clip, scale=0.34]{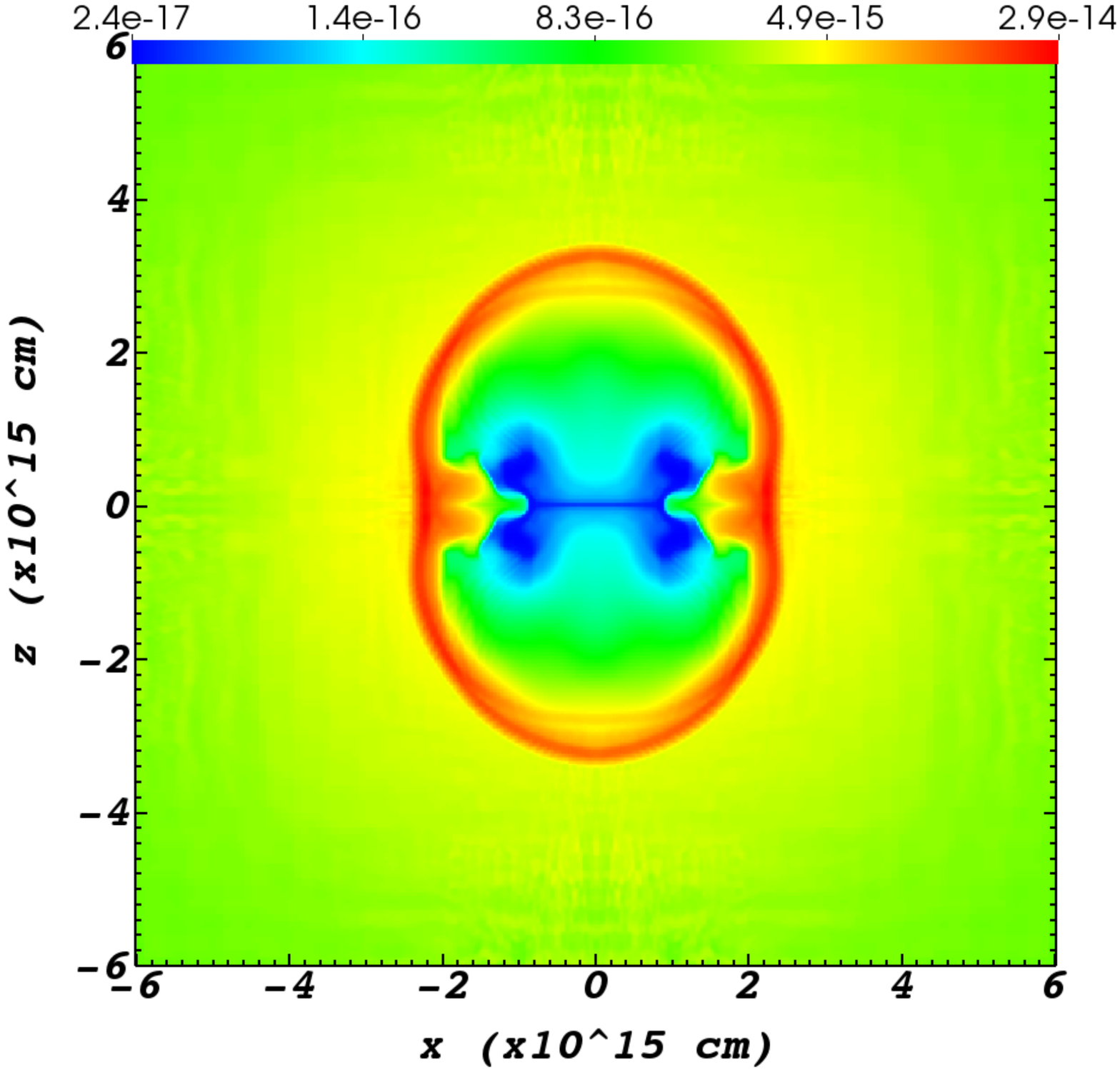} 
\end{tabular}
\caption{  Density maps in the meridional plane of simulations RR10, RR20, and RR50, from top to bottom, respectively. 
The time of the maps is $t=142 \days$, so they can be compared to the bottom panel of Fig. \ref{fig:HRCase1Evolution}. The middle panel is for the same physical parameters as that of figure Fig. \ref{fig:HRCase1Evolution}, but has a lower numerical resolution.  
}
  \label{fig:Angles}
    \end{figure}

 We conduct three simulations with the regular resolution (RR). Simulation RR20 has all the other parameters as in the HR simulation, including the half opening angle of the jets, while in simulations RR10 and RR50 we set  the jets' half opening angle to be $\alpha_{\rm j}=10^\circ$ and $\alpha_{\rm j}=50^\circ$, respectively. We present the density maps for these three simulations at $t=142 \days$ in Fig. \ref{fig:Angles}. Let us compare these  to the bottom panel of Fig. \ref{fig:HRCase1Evolution}.  

 We find that the regular resolution simulation (middle panel of Fig. \ref{fig:Angles}) gives the same basic bipolar structure as that of the high resolution simulation (bottom panel of Fig. \ref{fig:HRCase1Evolution}). The differences, as expected, are in the fine details of the flow, in particular near the symmetry axis, that the higher resolution simulations resolves better. This comparison assures us that the regular resolution is adequate for our goals.  

 We discuss now the role of he jets' half opening angle $\alpha_{\rm j}$. As expected, the narrower jet inflate a more elongated bipolar structure. Although the ram pressure of the jets in simulation RR10 ($\alpha_{\rm j}=10^\circ$) is four times that of the jets in simulation RR20 ($\alpha_{\rm j}=20^\circ$), the length of bipolar lobes (along the$z$ axis) is only $15\%$ larger in simulation RR10 than in simulation RR20. The width of the waist (in the equatorial plane) of simulation RR20 is $15\%$ larger than in simulation RR10. We conclude that our results are not sensitive to the jets' half opening angle as long as the jets are not very wide. 
We find that for the parameters we use here we lose the bipolar structure, of two bubbles with a narrow waist between them, at a half opening angle of $\alpha_{\rm j} \simeq 50^\circ$. We present this limiting case in the lower panel of Fig. \ref{fig:Angles}. We see that the waist is almost gone in the bipolar structure at this time of $t=142 \days$. The inflated bubbles form now an elliptical structure rather than a bipolar one.

\subsection{Other cases}
\label{subsec:OtherCases}

We conducted a simulation of a case where the jets are 25 times more energetic by having 25 more mass relative to the high-resolution simulation. This HE simulation is of a lower resolution than that of the HR simulation (Table \ref{Table:cases}). All other parameters are as in the HR simulation. Because this simulation is of lower resolution, we study in this case only the bipolar structure, and not the bubbles and the other inner regions. 
\begin{figure} [ht]
\centering
\begin{tabular}{cc}
\includegraphics[trim=0.0cm 6.0cm 0cm 5.2cm ,clip, scale=0.34]{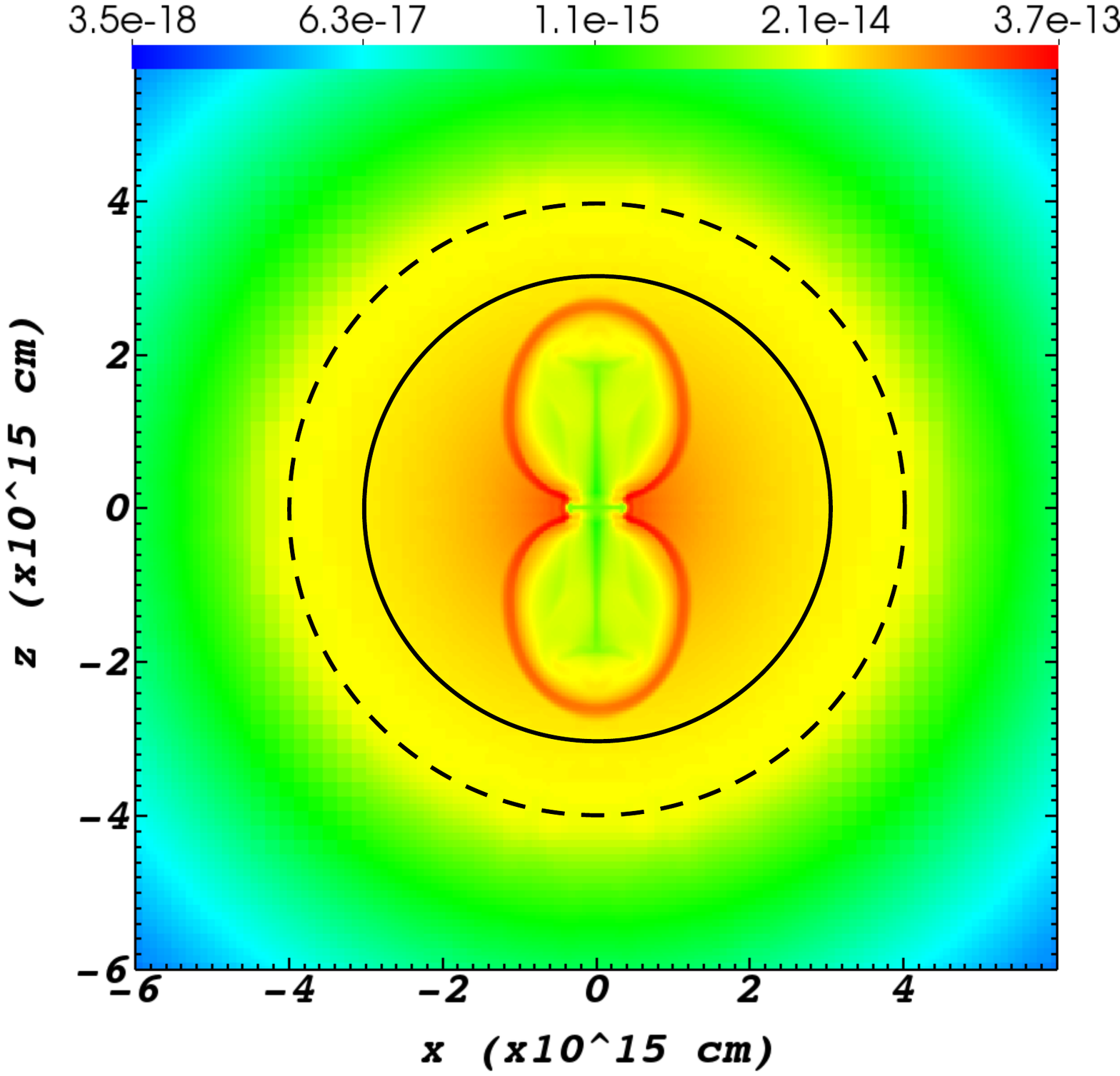} \\
\includegraphics[trim=0.0cm 6.0cm 0cm 5.2cm ,clip, scale=0.34]{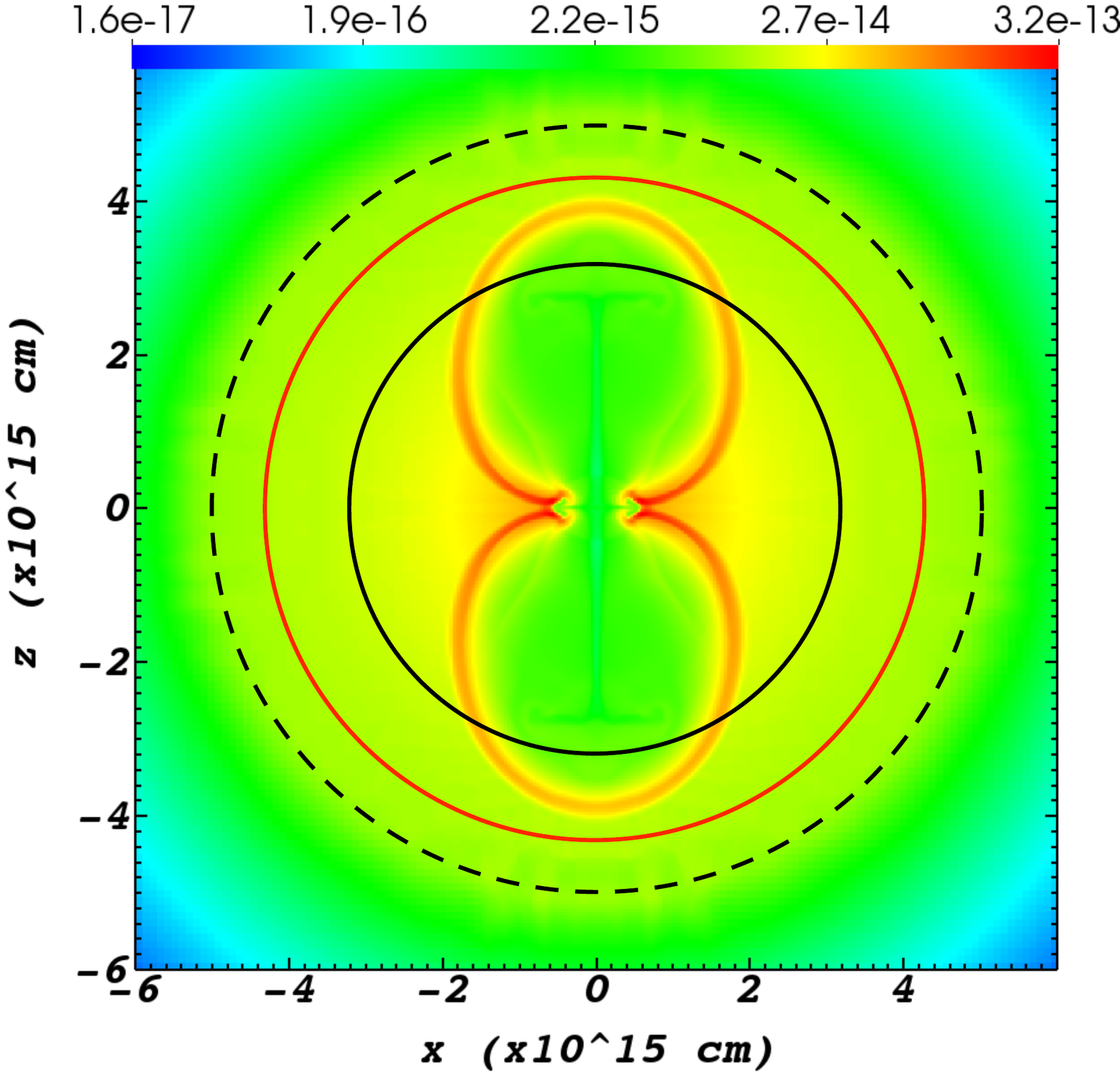}\\
\includegraphics[trim=0.0cm 6.0cm 0cm 5.2cm ,clip, scale=0.34]{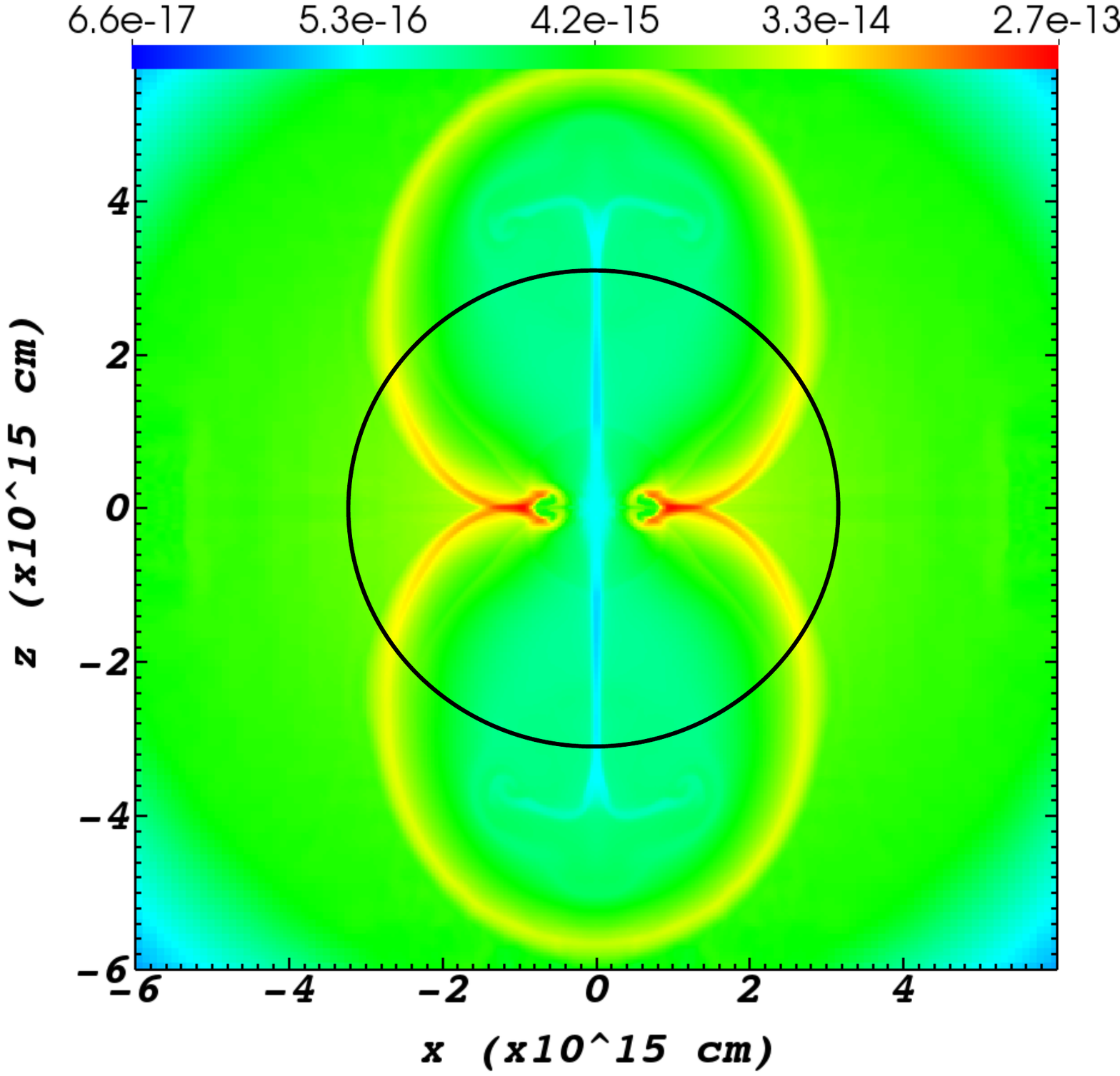} 
\end{tabular}
\caption{Similar to Fig. \ref{fig:HRCase1Evolution}, but for the high-energy (HE) simulation where the jets are 25 times more energetic by having 25 times more mass, and at three different times of $t=62$, 76, and 96~days, from top to bottom, respectively. 
}
  \label{fig:LR25energy}
    \end{figure}

As expected, the bipolar structure grows faster relative to the HR simulation. By the time the front of the dens shells break through the photosphere (on both sides), we clearly see that the density in the bubbles have much lower densities, by about an order of magnitude, relative to the densities in the equatorial plane at the same distances (middle panel of Fig. \ref{fig:LR25energy}). This implies that the photosphere will recede very rapidly within the bipolar structure (the bubbles), something that will lead to a rapid drop in the light curve for an equatorial observer (\citealt{KaplanSoker2020b, SokerKaplan2021}; section \ref{subsec:Evolution} above). 

\begin{figure}  [ht]
\centering
\begin{tabular}{cc}
\includegraphics[trim=0.0cm 6.0cm 0cm 5.2cm ,clip, scale=0.34]{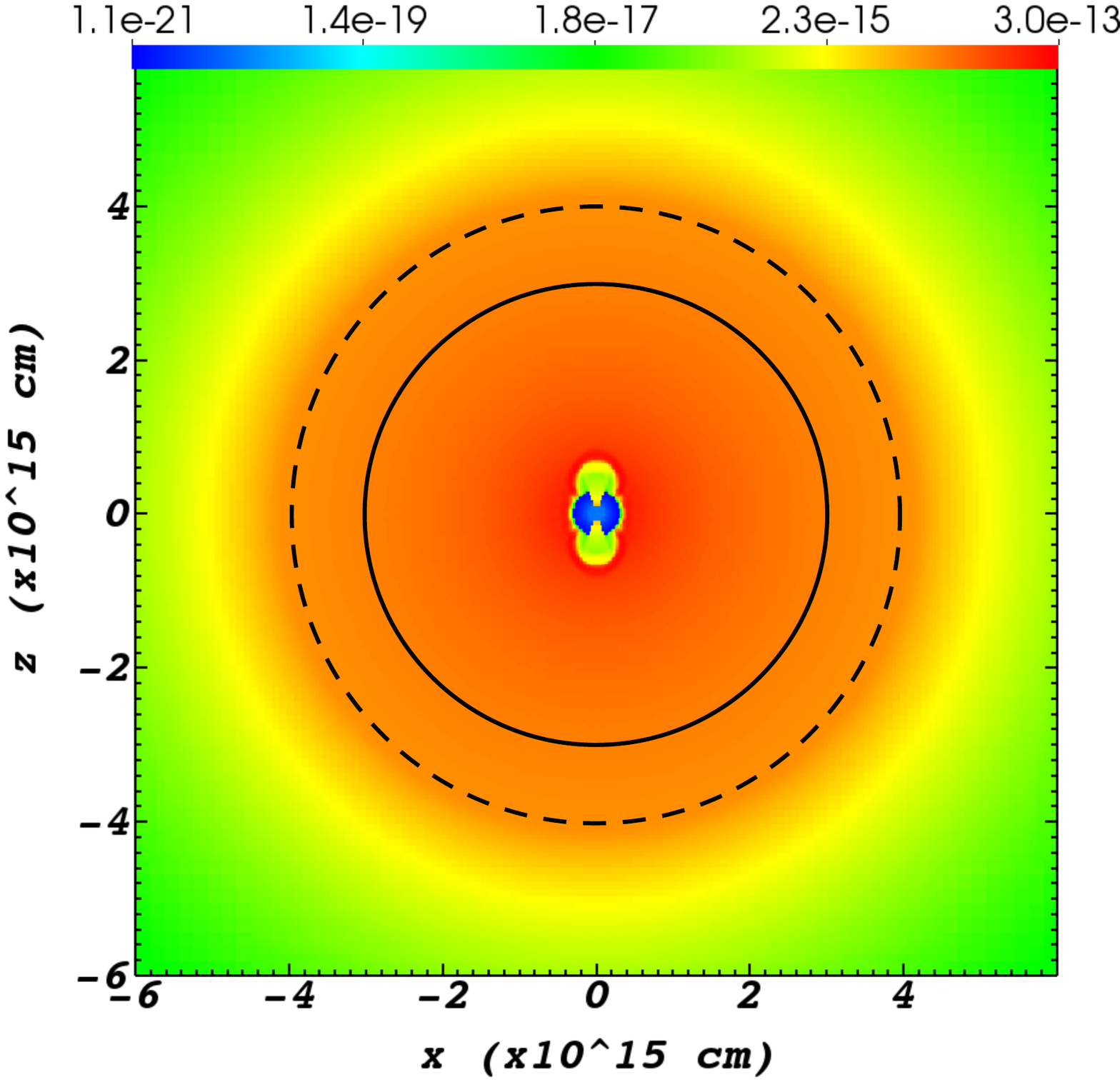} \\
\includegraphics[trim=0.0cm 6.0cm 0cm 5.2cm ,clip, scale=0.34]{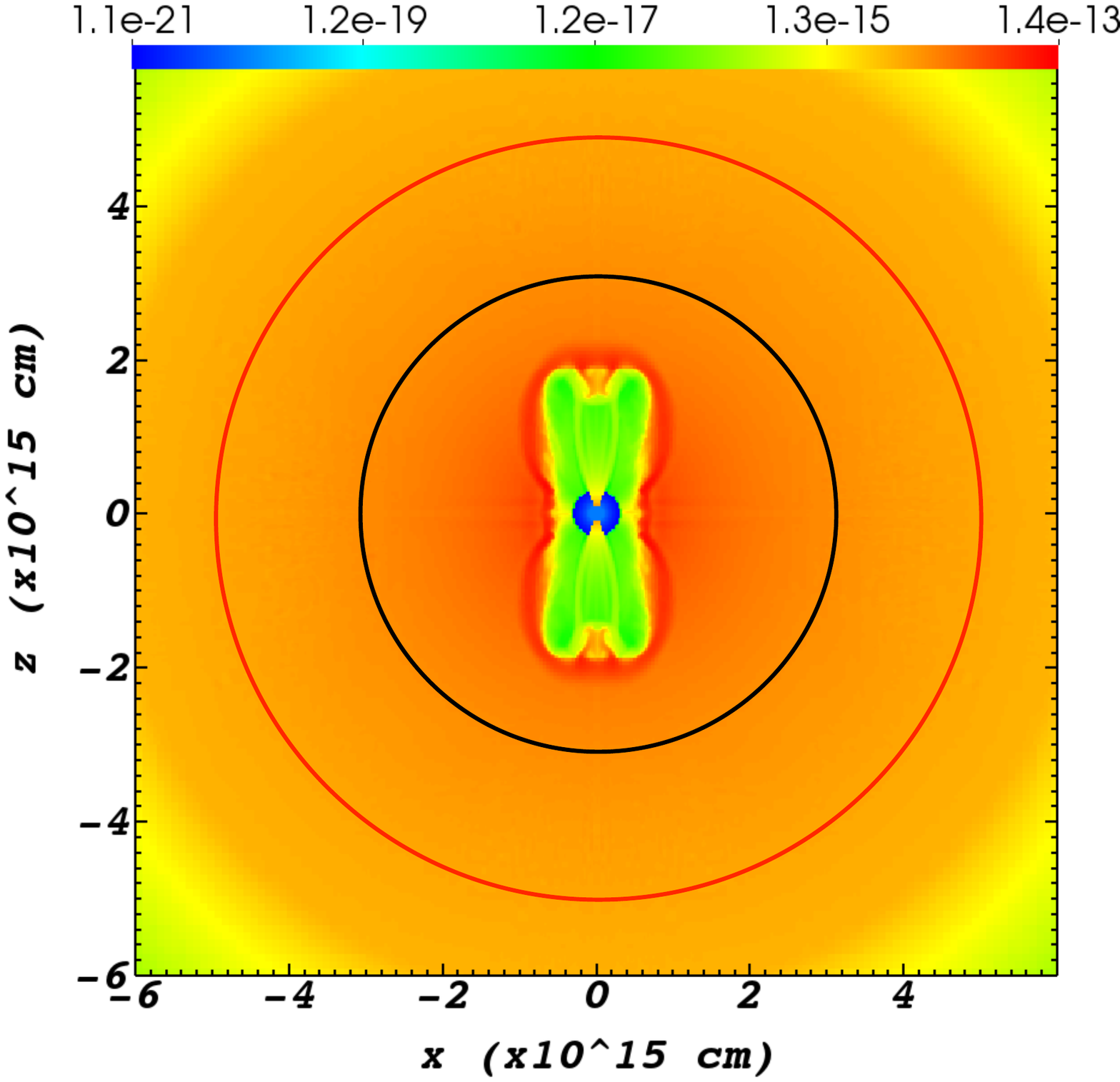}\\
\includegraphics[trim=0.0cm 6.0cm 0cm 5.2cm ,clip, scale=0.34]{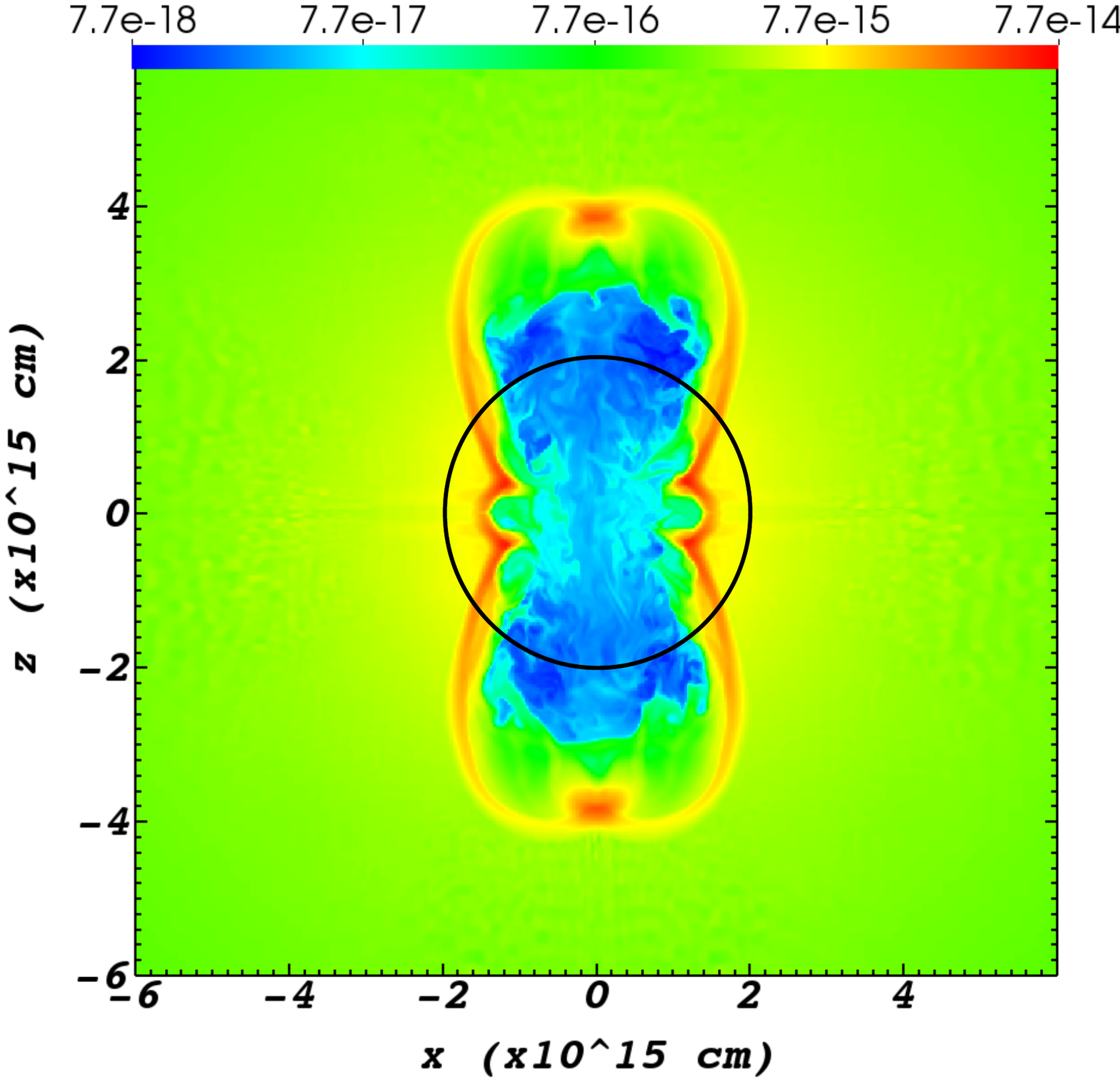} 
\end{tabular}
\caption{Similar to Fig. \ref{fig:HRCase1Evolution}, but for the long-activity (LA) simulation where the jets are active for 50 days. The times of the three frames are also as in Fig. \ref{fig:HRCase1Evolution}, $t=62$, $96$, and $142$~days.
}
  \label{fig:LA50days}
    \end{figure}

In Fig. \ref{fig:LA50days} we present the density maps at at three times for a long-activity (LA) simulation where the jets were active for $50~\days$, starting at $t_0=50 \days$, and the energy is as in the HR simulation. Comparing this figure to Fig. \ref{fig:HRCase1Evolution}, we immediately see that the dense shell is more elongated. In the HR simulation (Fig. \ref{fig:HRCase1Evolution}) the ratio of the length of one bubble (or the dense shell on one side) to its full width at maximum width at $t=142 \days$ is $\beta({\rm HR})=0.9$, while in the LA simulation (Fig. \ref{fig:LA50days}) this ratio is $\beta({\rm LA})=1.15$ at the same time. The same holds for the very low-density inner bubbles (deep blue), which are much more extended in the LA simulation. This might imply an even more abrupt drop in the light curve for an equatorial observer than we discussed above \citep{KaplanSoker2020b}. 

\begin{figure}  [ht]
\centering
\begin{tabular}{cc}
\includegraphics[trim=0.0cm 6.0cm 0cm 5.2cm ,clip, scale=0.34]{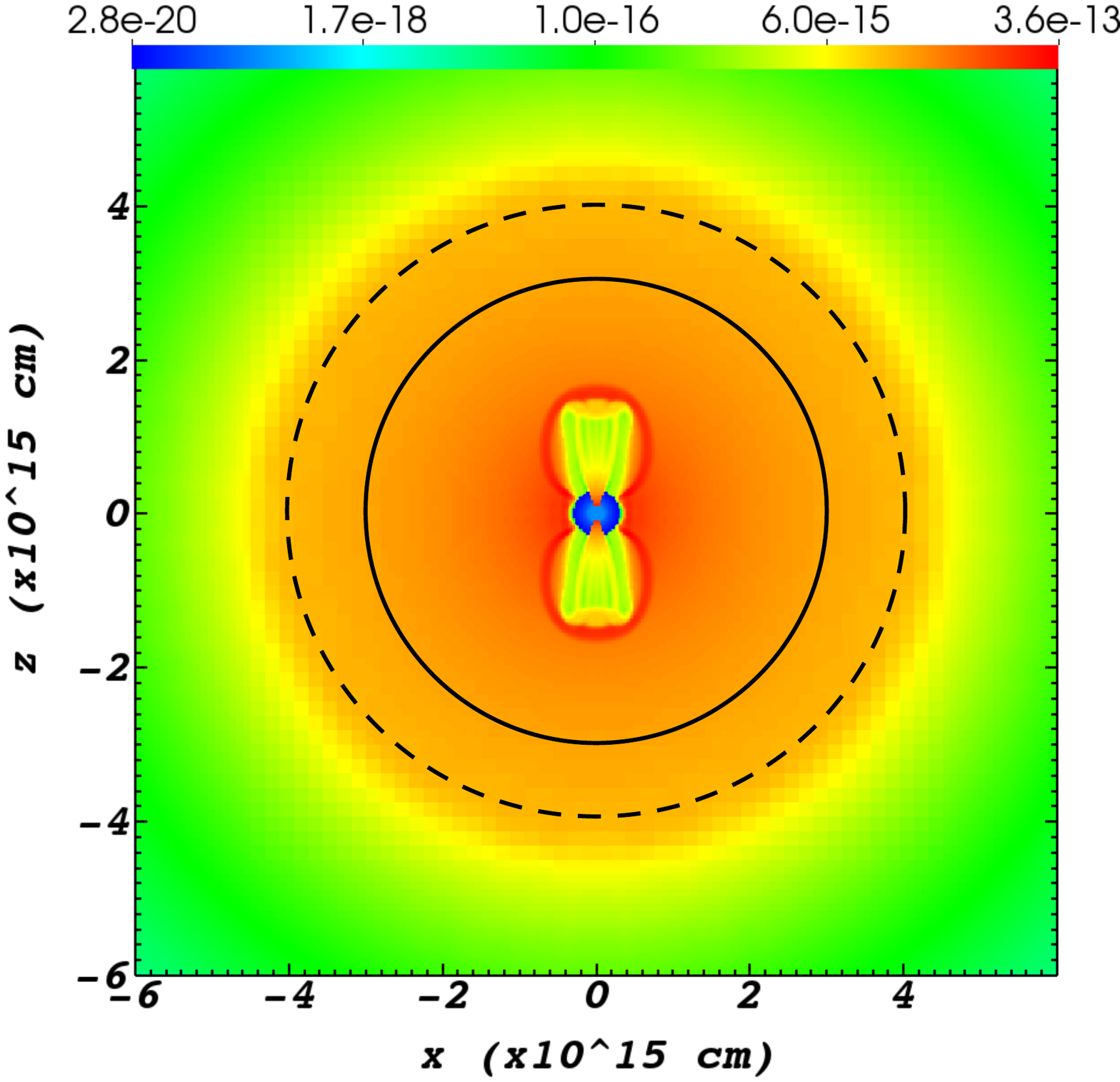} \\
\includegraphics[trim=0.0cm 6.0cm 0cm 5.2cm ,clip, scale=0.34]{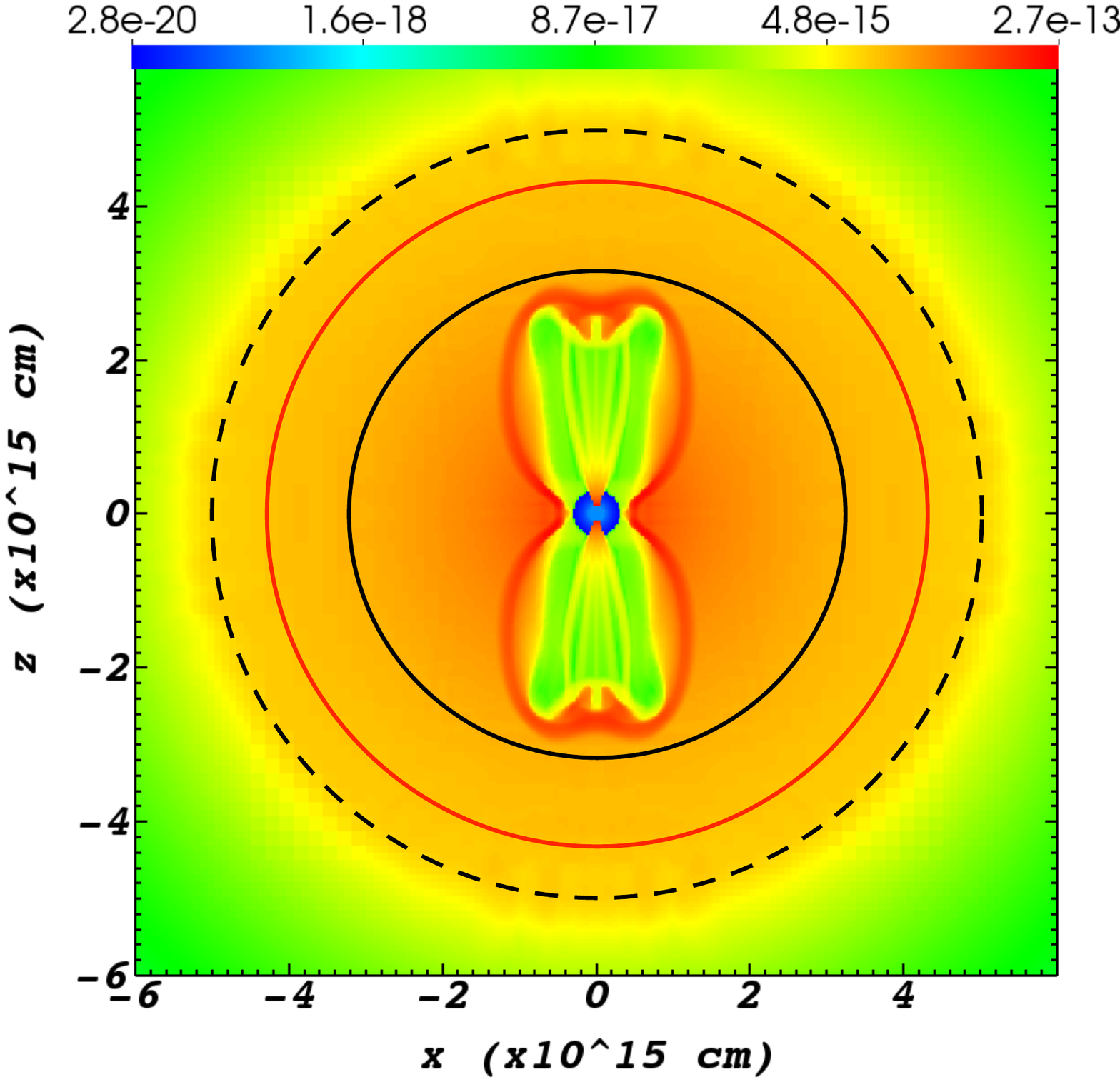}\\
\includegraphics[trim=0.0cm 6.0cm 0cm 5.2cm ,clip, scale=0.34]{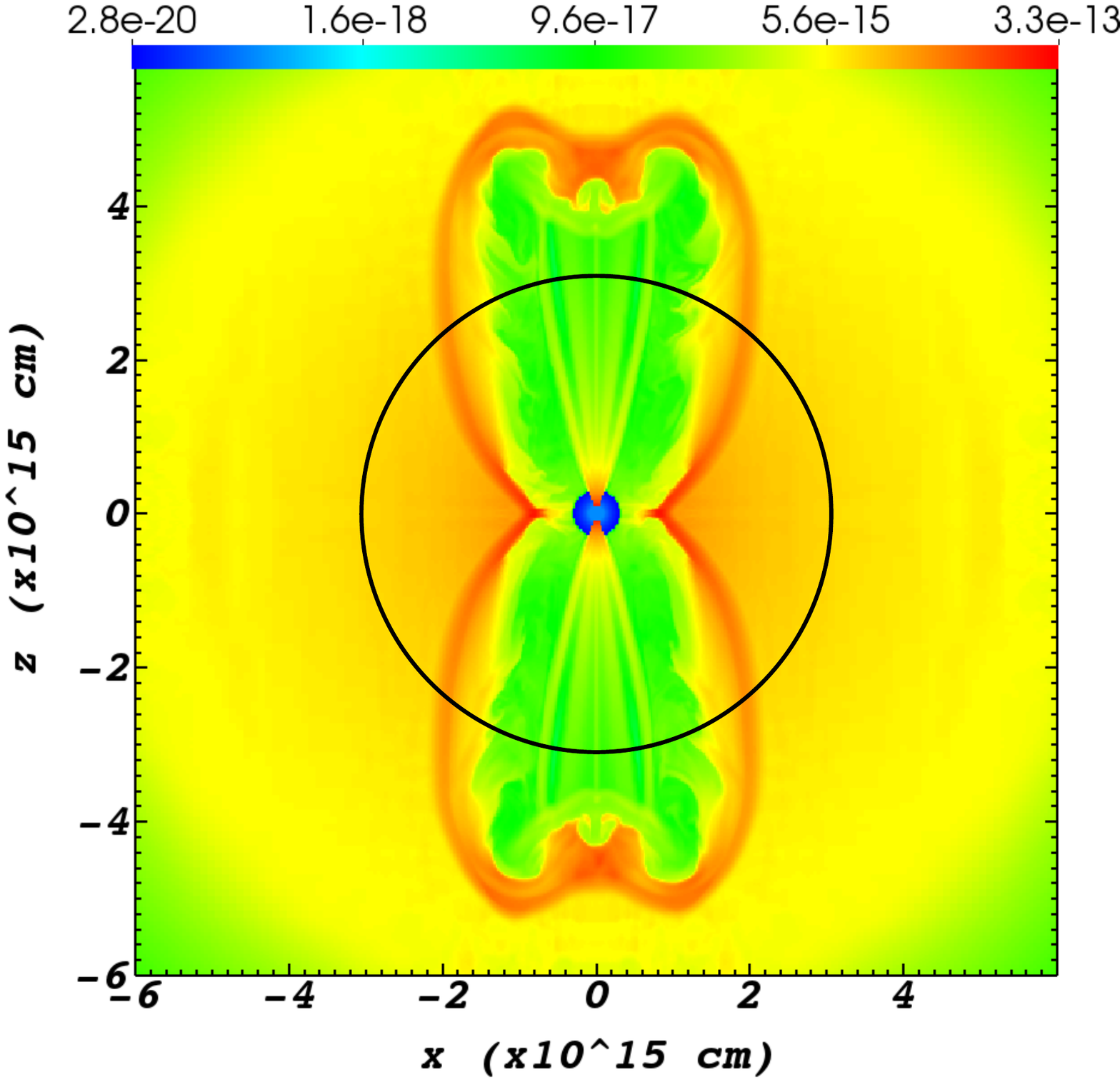} 
\end{tabular}
\caption{Similar to Fig. \ref{fig:LR25energy}, but for the long-activity high-energy (LAHE) simulation where the jets are active for 50 days and the energy is high (Table \ref{Table:cases}). The times of the three frames are also as in Fig. \ref{fig:LR25energy}, $t=62$, $76$, and $96$~days.
}
  \label{fig:LAHE50days}
    \end{figure}

In Fig. \ref{fig:LAHE50days} we present the density maps at at three times for a long-activity high-energy (LAHE) simulation where the jets were active for 50~days, starting at $t_0=50 \days$, and the energy is as in the HE simulation, i.e., 25 times the energy in the HR and LA simulations.  In the HE simulation (Fig. \ref{fig:LR25energy}) the ratio of the length of one bubble to its full width at maximum width at $t=96 \days$ is $\beta({\rm HE})=1.05$, while in the LAHE simulation (Fig. \ref{fig:LAHE50days}) this ratio is $\beta({\rm LAHE})=1.2$ at the same time.
The bubbles are of very low density, about an order of magnitude lower, relative to the regions outside the dense shells. As we commented above, this supports the toy model that \cite{KaplanSoker2020b} assumed, and therefore supports their suggestion that this might lead to an abrupt drop in the light curve for an equatorial observer when the photosphere enters the bubbles (lower panel of Fig. \ref{fig:LAHE50days}).    

 We note in the lower panel of Fig. \ref{fig:LAHE50days} the instabilities that develop in the inner boundary of the dense shell. These are the yellow zones that penetrate into the green region inside the lobes. In this simulation the process of jet-ejecta interaction is long-lived, and instabilities have time to develop.   

\begin{figure}
	\centering
\includegraphics[trim=1.7cm 5.9cm 0.0cm 3.8cm ,clip, scale=0.42]{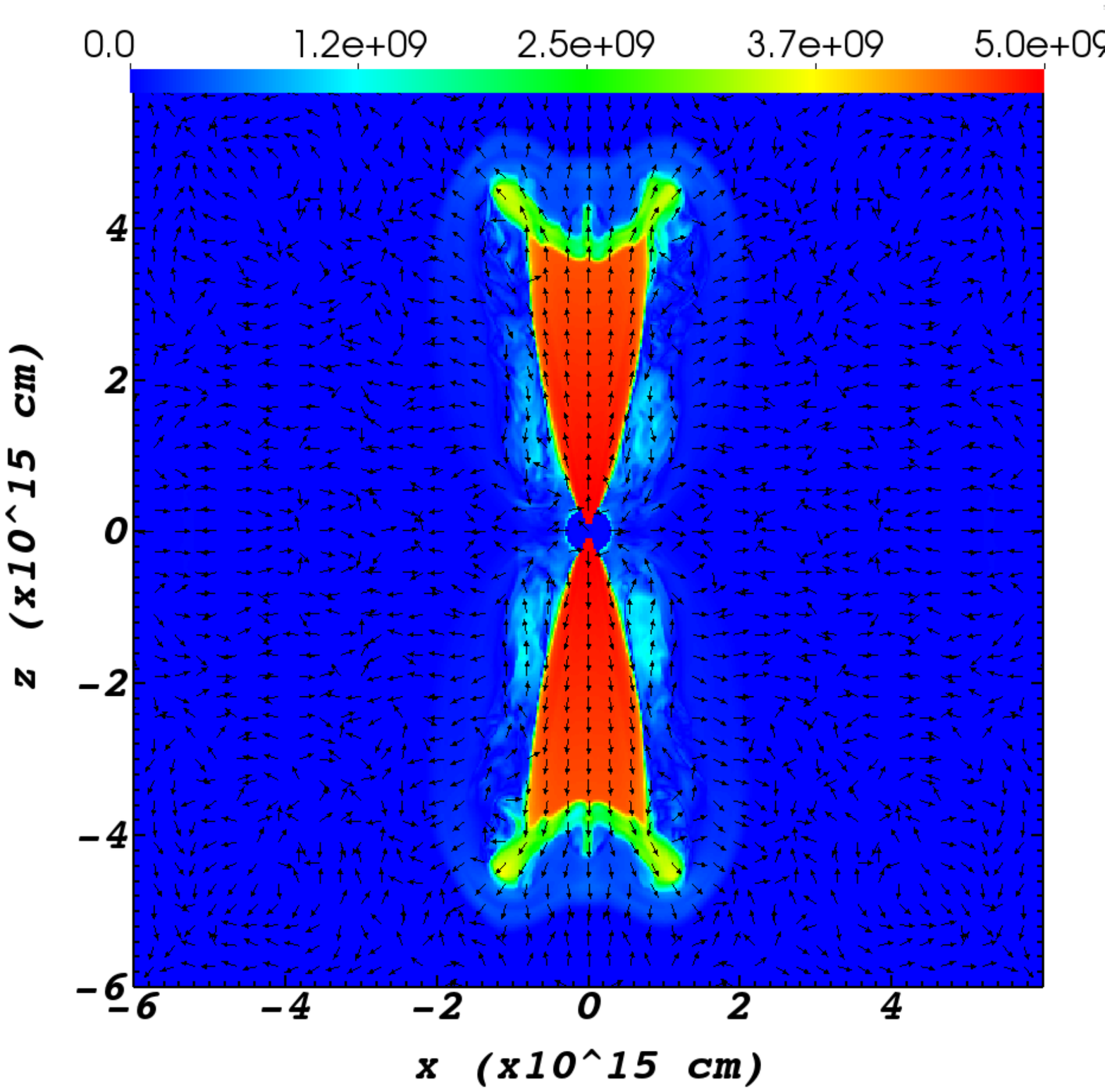}
	\caption{Velocity relative to the homologous expansion, $v_{\rm rel}$ from equation (\ref{eq:Vrel}), for the LAHE simulation. Colours indicate the magnitude of the velocity according to the colour-bar in $\cm \s^{-1}$. The arrows show only the direction of $v_{\rm rel}$. Note that the relative velocities in the deep-blue areas are very-very small (relative to the homologous expansion), and show numerical noise rather than flow structure.
	}
	\label{fig:VelocityMapLAHE}
\end{figure}

The energetic jets in the LAHE simulation, 
$E_{\rm 2j} = 0.2 E_{\rm SN}$, and their long activity time lead to an interaction that forms a deep along the symmetry axis, one at each side, in the outer boundary of each of the two dense shells. This is a feature we have obtained before when simulating jets in planetary nebulae (\citealt{AkashiSoker2016}). 
To further present the complicated flow structure, in Fig. \ref{fig:VelocityMapLAHE} we present the velocity map in the meridional plane of the LAHE simulation.

\subsection{Evolution of the bipolar shape}
\label{subsec:Shape}
 
\begin{figure}  [ht]
\centering
\begin{tabular}{cc}
\includegraphics[trim=0.7cm 7.0cm 0cm 6.2cm ,clip, scale=0.42]{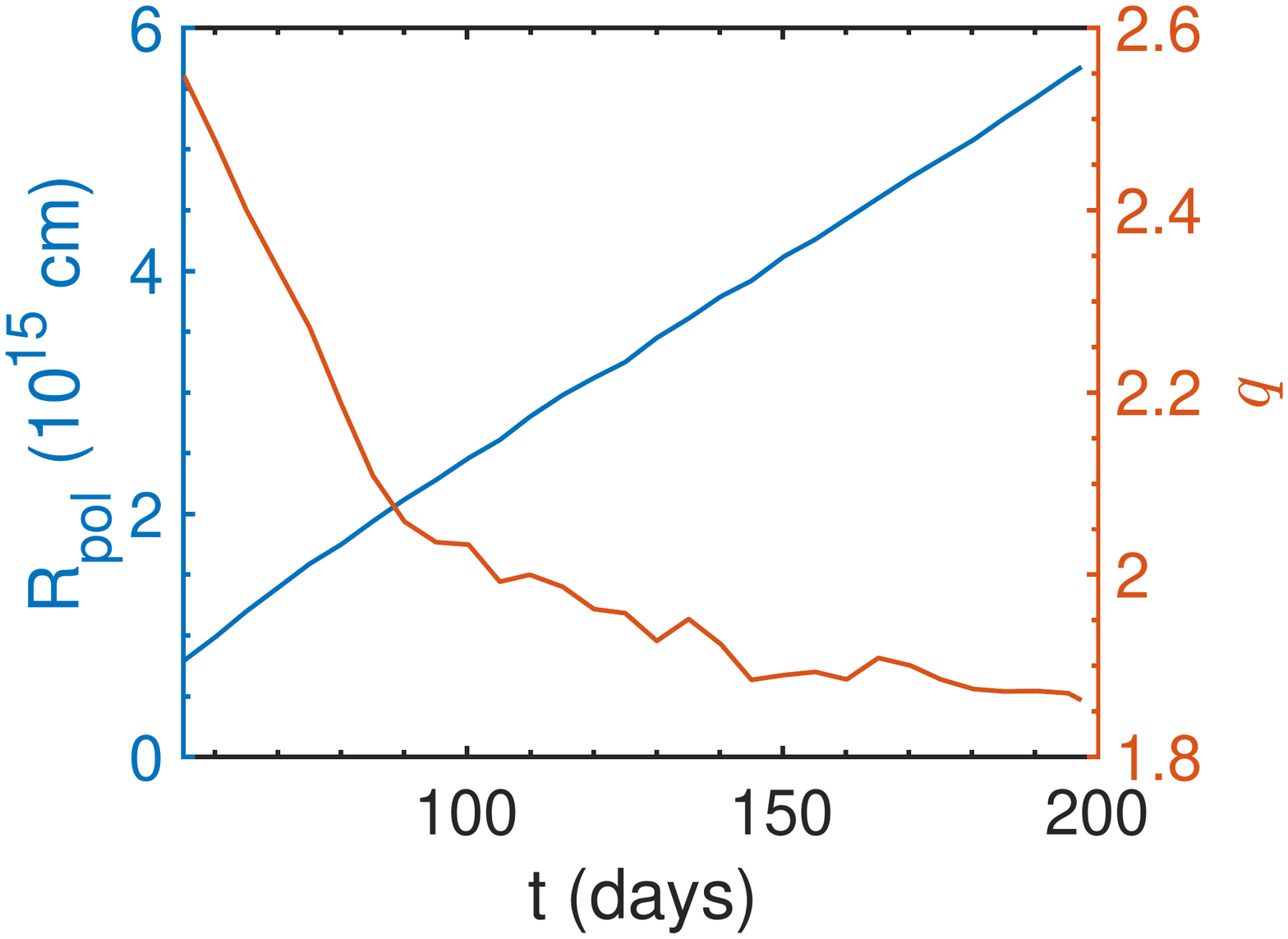} \\
\includegraphics[trim=0.7cm 7.0cm 0cm 7.2cm ,clip, scale=0.42]{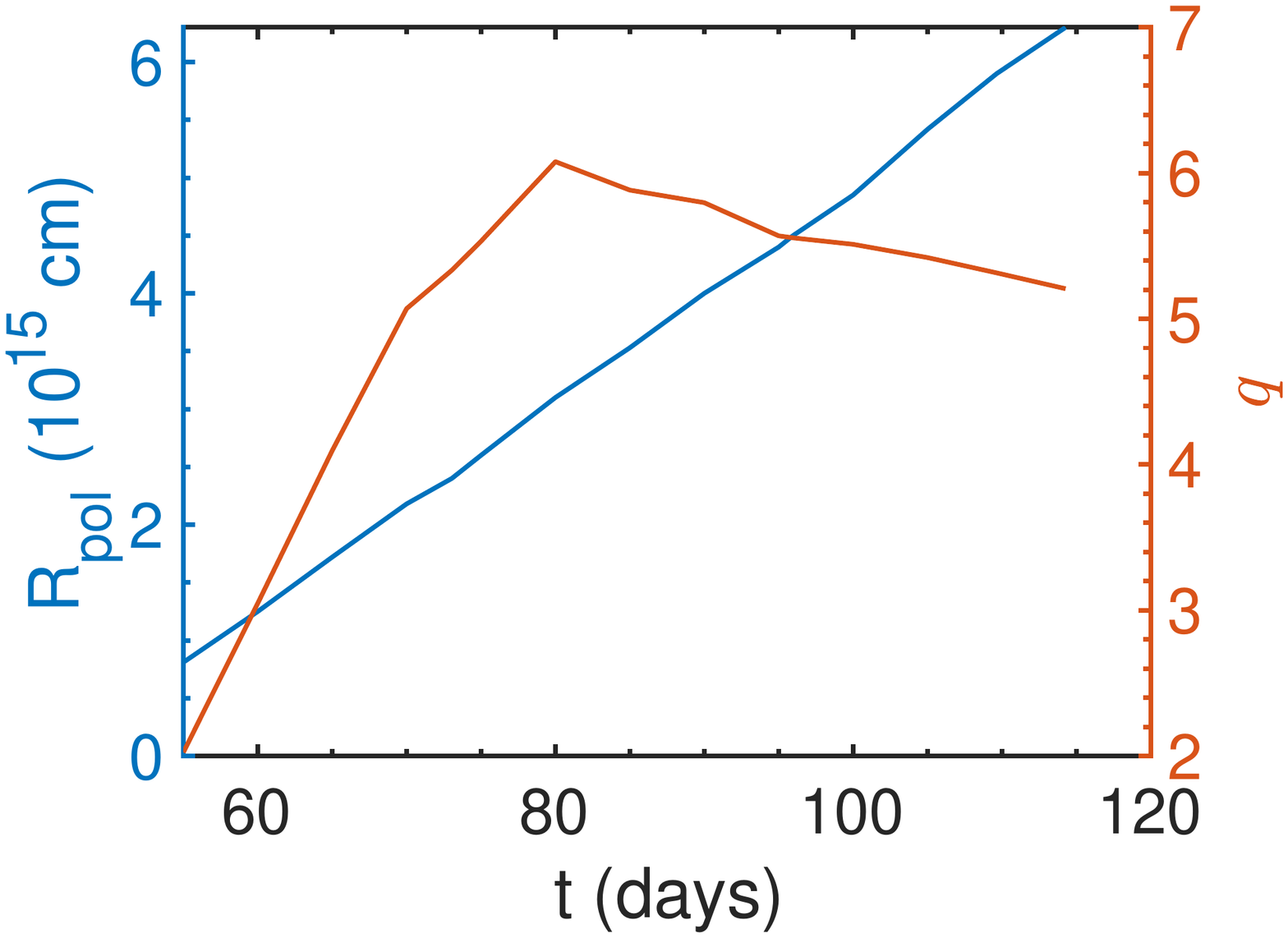} 
\end{tabular}
\caption{The evolution with time of $R_{\rm pol}$ and of $q \equiv R_{\rm pol}/R_{\rm equ}$, where $R_{\rm pol}$ is distance from center to the densest part in the dense shell along the polar axis and $R_{\rm equ}$ is this distance in the equatorial plane. In all cases we start the simulation at $t=50 \days$ after explosion. The upper panel presents this ratio for simulation RR20 where the jets are short-lived, and the lower panel shows the evolution for simulation LAHE where the jets are active for 50 days, i.e., from $t=50 \days$ to $t= 100 \days$.    
}
  \label{fig:NonSelf}
    \end{figure}

We examined the evolution of the ratio of polar to equatorial dimension, 
$q \equiv R_{\rm pol}/R_{\rm equ}$, where $R_{\rm pol}$ is distance from center to the densest part in the dense shell along the polar axis and $R_{\rm equ}$ is this distance in the equatorial plane. In Fig. \ref{fig:NonSelf} we present the evolution with time of $q$ and $R_{\rm pol}$ for simulations RR20 (a short jet activity) and LAHE (a long jet activity). 

We learn from Fig. \ref{fig:NonSelf} that in both cases the polar dimension expands more or less linearly with time. However, the ratio of polar dimension to equatorial dimension changes, and so the bipolar structure does not maintain the proportionality of its dimensions. In the case where the jet is short leaved (upper panel) the value of $q$ decreases, namely the bipolar structure becomes wider. In the case of  long-lived jets (lower panel) the value of $q$ increases until about $t=80 \days$, and then it decreases. We attribute the changing value of $q$ to the complicated flow structure near the equatorial plane, as we clearly see in Fig. \ref{fig:Pevolution} for simulation HR.  

\section{Discussion and Summary}
\label{sec:summary}

We conducted four 3D hydrodynamical simulations to study the interaction of late jets, either 50 days or 50 to 100 days after explosion, with the ejecta of a CCSN (table \ref{Table:cases}). We analysed the interaction flow in the HR simulation, where the jets are active for less than a day (Figs. \ref{fig:DensityMarks} - \ref{fig:HRCase1Evolution}). We mark the relevant morphological features on Fig. \ref{fig:DensityMarks}. 
In such a case of short-activity jets, the outcome looks like one explosion at each side of the equatorial plane (what \citealt{KaplanSoker2020a}  termed `min-explosion'). The morphology at late times is of an almost spherical bubble that is bounded by a dense-shell, one at each side of the equatorial plane (Fig. \ref{fig:HRCase1Evolution}). The same holds for the more energetic HE simulation with a short activity jets (Fig. \ref{fig:LR25energy}). 
In cases where the jets are active for a long time, the LA (fig. \ref{fig:LA50days}) and the LAHE (Fig. \ref{fig:LAHE50days}) simulations, the bubbles are more elongated. 
 For the parameters of this study, when the jets' half-opening angle is larger than about $50^\circ$, the two bubbles merge to form one elliptical shell (Fig. \ref{fig:Angles}). 
 
In Figs. \ref{fig:HRCase1Evolution}, and  \ref{fig:LR25energy} - \ref{fig:LAHE50days} we presented also the crude photosphere location. These figures represent the qualitative results that the bubbles break out from the photosphere at much earlier times than the regions at the same radii near the equatorial plane. In section \ref{subsec:Evolution} we discussed two implications of this. The first is a possible peak in the light curve at late times \citep{KaplanSoker2020a}. For a polar observer this peak might be blue in some cases 
\citep{KaplanSoker2020b}.
The second implication is that after the photosphere rapidly recedes inside the bubble (because of its low density), an equatorial observer does no see the polar photosphere any more \citep{KaplanSoker2020b}. This might lead to a rapid light-curve drop for an equatorial observer \citep{KaplanSoker2020b}. 
Our results support the toy models that  \cite{KaplanSoker2020a} and \cite{KaplanSoker2020b} assumed, and therefore support their conclusions regarding possible late peaks in the light curve and rare cases of rapid drop in luminosity for an equatorial observer. 
 
Although we simulated late jets from the center, our results on the formation of bipolar hot and low-density bubbles have wider implications.
Our results can be extended to jets that explode the star. Namely, very energetic jets at the explosion itself. If the pre-collapse core has only a slow rotation, the explosion will be by jittering jets (assuming the jittering jets explosion mechanism). The jets in each jets-launching episode carry a small fraction of the total explosion energy (section \ref{sec:intro}). However, if the core has a large amount of angular momentum the jets might maintain a constant axis and lead to a super-energetic CCSN \citep{Gilkisetal2016Super}. Energetic jets that maintain a constant axis will form a bipolar structure with a similar morphology to what we have obtained in this study, but that extends to a large distance and occupies a large volume out of that of the ejecta. In a short time, days after explosion, the photosphere might be inside the hot-tenuous bubbles, leading to blue emission that drops within days. 

 Our results might be related to some other astrophysical objects. At this stage we only point these possible similarities to motivate further simulations of jets in these objects (e.g.,  \citealt{LopezCamaraetal2019, LopezCamaraetal2020}).  A similar  process of a strong blue emission followed by a rapid drop might take place when the jets that explode the star are of a NS companion that merges with the core, the so called common envelope jets supernova (CEJSN). \cite{Soker2019AT2018Cow} suggested that the fast blue optical transients (FBOT) AT2018cow \citep{Prenticeetal2018} was a CEJSN. 
In the specific CEJSN scenario that \cite{Soker2019AT2018Cow} proposed for AT2018cow, the jets clear the polar directions of the giant envelope before the NS companion launches the jets that explodes the star as it accretes mass from the core. They termed this the polar CEJSN scenario. 
The basic process in the polar CEJSN is the formation of two opposite bubbles in a bipolar structure. 
 Our results might have relation to the morphological feature that \cite{Soker2019AT2018Cow} proposed, but the later has much large bubbles, and for that requires a new set of simulations.  
 If this similarity holds,  we suggest here the CEJSN scenario to two other FBOTs, ZTF18abvkwla \citep{Hoetal2020} and CSS161010 \citep{Coppejansetal2020}. 

We also note that a very close, about $1-5 R_\odot$, NS companion to an exploding stripped-envelope (Type Ib or Ic) CCSN might launch weak jets that form an elongated structure in the inner regions of the ejecta \citep{Soker2020NS2}, similar in some aspects to the structures we find here,  as we show in a very recent study \citep{AkashiSoker2020}.

\section*{Acknowledgments}

 We thank an anonymous referee for very useful comments and suggestions.  This research was supported by a grant from the Israel Science Foundation (420/16 and 769/20) and a grant from the Asher Space Research Fund at the Technion.

\textbf{Data availability}
The data underlying this article will be shared on reasonable request to the corresponding author.  

\label{lastpage}

\begin{thebibliography}{}

\bibitem[Akashi \& Soker(2016)]{AkashiSoker2016} Akashi, M., \& Soker, N.\ 2016, \mnras, 462, 206


\bibitem[Akashi \& Soker(2020)]{AkashiSoker2020} Akashi, M. \& Soker, N.\ 2020, \apj, 901, 53. doi:10.3847/1538-4357/abad35

\bibitem[Aloy et al.(2000)]{Aloyetal2000} Aloy M. A., Muller E., Ibanez J.~M., Marti, J.~M., \& MacFadyen A.\ 2000, \apj, 531, L119
	
\bibitem[Bear et al.(2017)]{Bearetal2017} Bear, E., Grichener, A., \& Soker, N.\ 2017, \mnras, 472, 1770
	
\bibitem[Bear, \& Soker(2018)]{BearSoker2018} Bear, E., \& Soker, N.\ 2018, \mnras, 478, 682
	
\bibitem[Bethe \& Wilson(1985)]{BetheWilson1985} Bethe, H.~A., \& Wilson, J.~R.\ 1985, \apj, 295, 14

\bibitem[Bose et al.(2019)]{Boseetal2019} Bose S., et al., 2019, ApJ, 873, L3

\bibitem[Bromberg \& Tchekhovskoy(2016)]{BrombergTchekhovskoy2016} Bromberg, O., \& Tchekhovskoy, A.\ 2016, \mnras, 456, 1739
	
\bibitem[Bromberg et al.(2014)]{Bromberg_jet} Bromberg, O., Granot, J., Lyubarsky, Y., et al.\ 2014, \mnras, 443, 1532.
	
\bibitem[Bruenn et al.(2016)]{Bruennetal2016} Bruenn, S.~W., Lentz, E.~J., Hix, W.~R., et al.\ 2016, \apj, 818, 123

\bibitem[Bugli et al.(2020)]{Buglietal2020} Bugli, M., Guilet, J., Obergaulinger, M., Cerd{\'a}-Dur{\'a}n, P., \& Aloy, M.~A., \ 2020, \mnras, 492, 58

\bibitem[Burrows et al.(2007)]{Burrows2007} Burrows, A., Dessart, L., Livne, E., Ott, C.~D., \& Murphy, J.\ 2007, \apj, 664, 416

\bibitem[Burrows et al.(2018)]{Burrowsetal2018} Burrows, A., Vartanyan, D., Dolence, J.~C., Skinner, M.~A., \& Radice, D.\ 2018, \ssr, 214, 33

\bibitem[Casanova et al.(2020)]{Casanovaetal2020} Casanova, J., Endeve, E., Lentz, E.~J., Messer, O.~E.~B., Hix, W.~R., Harris, J.~A., \& Bruenn S.~W.,\ 2020, arXiv e-prints, arXiv:2004.02055
  
\bibitem[Coppejans et al.(2020)]{Coppejansetal2020} Coppejans, D.~L., Margutti, R., Terreran, G., et al.\ 2020, \apjl, 895, L23

\bibitem[Couch \& Ott(2013)]{CouchOtt2013} Couch, S.~M., \& Ott, C.~D.\ 2013, \apjl, 778, L7

\bibitem[Couch et al.(2020)]{Couchetal2020} Couch, S.~M., Warren, M.~L., \& O'Connor, E.~P.\ 2020, \apj, 890, 127

\bibitem[Delfan Azari et al.(2020)]{DelfanAzarietal2020} Delfan Azari, M., Yamada, S., Morinaga, T., et al.\ 2020, \prd, 101, 023018

\bibitem[Feng et al.(2018)]{Fengetal2018} Feng, E.-H., Shen, R.-F., \& Lin, W.-P.\ 2018, \apj, 867, 130

\bibitem[Fryxell et al.(2000)]{Fryxell2000} Fryxell, B., Olson, K., Ricker, P., et al.\ 2000, \apjs, 131, 273
	
\bibitem[Garc{\'{\i}}a et al.(2017)]{Garciaetal2017} Garc{\'{\i}}a, F., Su{\'a}rez, A.~E., Miceli, M., Bocchino, F., Combi, J.~A., Orlando, S., \& Sasaki, M.\ 2017, \aap, 604, L5

\bibitem[Gilkis(2018)]{Gilkis2018} Gilkis, A.\ 2018, \mnras, 474, 2419

\bibitem[Gilkis \& Soker(2014)]{GilkisSoker2014}  Gilkis, A., \& Soker, N.\ 2014, \mnras, 439, 4011

\bibitem[Gilkis \& Soker(2015)]{GilkisSoker2015}  Gilkis, A., \& Soker, N.\ 2015, \apj, 806, 2

\bibitem[Gilkis et al.(2016)]{Gilkisetal2016Super} Gilkis, A., Soker, N., \& Papish, O.\ 2016, \apj, 826, 178

\bibitem[Gofman et al.(2020)]{Gofmanetal2020} Gofman, R.~A., Gluck, N., \& Soker, N.\ 2020, \mnras, 494, 5230

\bibitem[Gonz{\'a}lez-Casanova et al.(2014)]{Gonzalezetal2014} Gonz{\'a}lez-Casanova, D.~F., De Colle, F., Ramirez-Ruiz, E., \& Lopez, L.~A.\ 2014, \apjl, 781, L26
	
\bibitem[Grichener, \& Soker(2017)]{GrichenerSoker2017} Grichener, A., \& Soker, N.\ 2017, \mnras, 468, 1226

\bibitem[Ho et al.(2020)]{Hoetal2020} Ho, A.~Y.~Q., Perley, D.~A., Kulkarni, S.~R., et al.\ 2020, \apj, 895, 49

\bibitem[Inserra et al.(2016)]{Inserraetal2016}  Inserra, C., Bulla, M., Sim, S.~A., \& Smartt, S.~J.\ 2016, \apj, 831, 79

\bibitem[Iwakami et al.(2020)]{Iwakamietal2020} Iwakami, W., Okawa, H., Nagakura, H., et al.\ 2020, arXiv e-prints, arXiv:2004.02091

\bibitem[Janka et al.(2016)]{Jankaetal2016} Janka, H.-T., Melson, T., \& Summa, A.\ 2016, Annual Review of Nuclear and Particle Science, 66, 341

\bibitem[Kaplan \& Soker(2020a)]{KaplanSoker2020a} Kaplan, N., \& Soker, N.\ 2020a, \mnras, 492, 3013

\bibitem[Kaplan \& Soker(2020b)]{KaplanSoker2020b} Kaplan, N., \& Soker, N.\ 2020b,  \mnras, 494, 5909. doi:10.1093/mnras/staa1201

\bibitem[Kazeroni \& Abdikamalov(2020)]{KazeroniAbdikamalov2020} Kazeroni, R., \& Abdikamalov, E.\ 2020, \mnras, doi:10.1093/mnras/staa944

\bibitem[Khokhlov et al.(1999)]{Khokhlovetal1999} Khokhlov, A.~M., H{\"o}flich, P.~A., Oran, E.~S., et al.\ 1999, \apjl, 524, L107
	
\bibitem[Kuroda et al.(2020)]{Kurodaetal2020} Kuroda, T., Arcones, A., Takiwaki, T., \& Kotake, K.,\ 2020, arXiv e-prints, arXiv:2003.02004

\bibitem[Lazzati et al.(2012)]{Lazzati2012} Lazzati, D., Morsony, B.~J., Blackwell, C.~H., \& Begelman, M.~C.\ 2012, \apj, 750, 68

\bibitem[Lopez \& Fesen(2018)]{LopezFesen2018} Lopez, L.~A., \& Fesen, R.~A.\ 2018, \ssr, 214, \#44

\bibitem[L{\'o}pez-C{\'a}mara et al.(2019)]{LopezCamaraetal2019} L{\'o}pez-C{\'a}mara, D., De Colle, F., \& Moreno M{\'e}ndez, E.\ 2019, \mnras, 482, 3646. doi:10.1093/mnras/sty2959

	
\bibitem[L{\'o}pez-C{\'a}mara et al.(2016)]{LopezCamaraetal2016}  L{\'o}pez-C{\'a}mara, D., Lazzati, D., \& Morsony, B.~J.\ 2016, \apj, 826, 180

\bibitem[L{\'o}pez-C{\'a}mara et al.(2020)]{LopezCamaraetal2020} L{\'o}pez-C{\'a}mara, D., Moreno M{\'e}ndez, E., \& De Colle, F.\ 2020, \mnras, 497, 2057. doi:10.1093/mnras/staa1983

	
\bibitem[L{\'o}pez-C{\'a}mara et al.(2014)]{LopezCamaraetal2014}  L{\'o}pez-C{\'a}mara, D., Morsony, B.~J., \& Lazzati, D.\ 2014, \mnras, 442, 2202

\bibitem[Mabanta et al.(2019)]{Mabantaetal2019} Mabanta, Q.~A., Murphy, J.~W., \& Dolence, J.~C.\ 2019, \apj, 887, 43

\bibitem[Maeda et al.(2012)]{Maedaetal2012} Maeda, K., Moriya, T., Kawabata, K., et al.\ 2012, \memsai, 83, 264
	
\bibitem[Margutti et al.(2014)]{Marguttietal2014} Margutti, R., Milisavljevic, D., Soderberg, A.~M., et al.\ 2014, \apj, 797, 107
	
\bibitem[Mauerhan et al.(2017)]{Mauerhanetal2017} Mauerhan, J.~C., Van Dyk, S.~D., Johansson, J., Hu, M., Fox, O.~D., Wang, L., Graham, M.~L., Filippenko, A.~V., \& Shivvers, I.\ 2017, \apj, 834, 118
	
\bibitem[Maund et al.(2007)]{Maundetal2007} Maund, J.~R., Wheeler, J.~C., Patat, F.,  Baade, D., Wang, L., H\"oflich, P.\ 2007, \mnras, 381, 201

\bibitem[Milisavljevic et al.(2013)]{Milisavljevic2013} Milisavljevic, D., Soderberg, A.~M., Margutti, R., et al.\ 2013, \apjl, 770, LL38
	
\bibitem[M{\"o}sta et al.(2014)]{Mostaetal2014} M{\"o}sta, P., Richers, S., Ott, C.~D., et al.\ 2014, \apjl, 785, L29

\bibitem[M{\"u}ller(2016)]{Muller2016} M{\"u}ller, B.\ 2016, \pasa, 33, e048

\bibitem[M{\"u}ller et al.(2019)]{Mulleretal2019Jittering} M{\"u}ller, B., Tauris, T.~M., Heger, A., et al.\ 2019, \mnras, 484, 3307

\bibitem[Nagakura et al(2011)]{Nagakuraetal2011} Nagakura H., Ito H., Kiuchi K., \& Yamada S.,\ 2011, ApJ, 731, 80

\bibitem[Nishimura et al.(2017)]{Nishimuraetal2017} Nishimura, N., Sawai, H., Takiwaki, T., Yamada, S., \& Thielemann, F.-K.\ 2017, \apjl, 836, L21
	
\bibitem[Orlando et al.(2016)]{Orlandoetal2016} Orlando S., Miceli M., Pumo M.~L., Bocchino F., 2016, ApJ, 822, 22

\bibitem[Papish \& Soker(2011)]{PapishSoker2011} Papish, O., \& Soker, N.\ 2011, \mnras, 416, 1697

\bibitem[Papish \& Soker(2014)]{PapishSoker2014} Papish, O., \& Soker, N.\ 2014, \mnras, 443, 664 

\bibitem[Powell \& M{\"u}ller(2020)]{PowellMuller2020} Powell, J., \& M{\"u}ller, B.\ 2020, \mnras, 494, 4665

\bibitem[Prentice et al.(2018)]{Prenticeetal2018} Prentice, S.~J., Maguire, K., Smartt, S.~J., et al.\ 2018, \apjl, 865, L3

\bibitem[Quataert et al.(2019)]{Quataertetal2019} Quataert, E., Lecoanet, D., \& Coughlin, E.~R.\ 2019, \mnras, 485, L83

\bibitem[Sawada, \& Maeda(2019)]{SawadaMaeda2019} Sawada, R., \& Maeda, K.\ 2019, \apj, 886, 47

\bibitem[Soker(2010)]{Soker2010} Soker, N.\ 2010, \mnras, 401, 2793

\bibitem[Soker(2018)]{Soker2018KeyRoleB} Soker, N.\ 2018, arXiv:1805.03447

\bibitem[Soker(2019a)]{Soker2019SASI} Soker, N.\ 2019a, Research in Astronomy and Astrophysics,  19, 095

\bibitem[Soker(2019b)]{Soker2019JitSim} Soker, N.\ 2019b, arXiv e-prints, arXiv:1907.13312

\bibitem[Soker(2020)]{Soker2020NS2} Soker, N.\ 2020, arXiv e-prints, arXiv:2005.07645

\bibitem[Soker \& Kaplan(2021)]{SokerKaplan2021} Soker, N. \& Kaplan, N.\ 2021, arXiv:2007.14021

\bibitem[Soker et al.(2019)]{Soker2019AT2018Cow} Soker, N., Grichener, A., \& Gilkis, A.\ 2019, \mnras, 484, 4972

\bibitem[Stockinger et al.(2020)]{Stockingeretal2020} Stockinger, G., Janka, H.-T., Kresse, D., et al.\ 2020,  \mnras, 496, 2039

\bibitem[Suzuki \& Maeda(2019)]{SuzukiMaeda2019} Suzuki, A., \& Maeda, K.\ 2019, \apj, 880, 150

\bibitem[Takiwaki \& Kotake(2011)]{TakiwakiKotake2011} Takiwaki, T., \& Kotake, K.\ 2011, \apj, 743, 30
	
\bibitem[Wang et al.(2001)]{Wangetal2001} Wang, L., Howell, D.~A., H{\"o}flich, P., \& Wheeler, J.~C.\ 2001, \apj, 550, 1030
		

\end{thebibliography}
\end{document}